\documentclass[11pt,a4paper]{article}

\usepackage[utf8]{inputenc}
\usepackage[T1]{fontenc}
\usepackage{lmodern}
\usepackage[margin=1in]{geometry}
\usepackage{amsmath,amssymb,amsfonts}
\usepackage{graphicx}
\usepackage{booktabs}
\usepackage{tabularx}
\usepackage{longtable}
\usepackage{array}
\usepackage{multirow}
\usepackage{xcolor}
\usepackage{listings}
\usepackage{enumitem}
\usepackage{float}
\usepackage{caption}
\usepackage{subcaption}
\usepackage{natbib}
\usepackage{url}
\usepackage{fancyhdr}
\usepackage{titlesec}
\usepackage{tikz}
\usetikzlibrary{positioning,fit,backgrounds,arrows.meta,decorations.pathreplacing,calc}
\usepackage{hyperref}
\usepackage{cleveref}

\hypersetup{
    colorlinks=true,
    linkcolor=blue,
    filecolor=magenta,
    urlcolor=cyan,
    citecolor=blue,
    breaklinks=true,
    bookmarksnumbered=true,
}

\fancypagestyle{plain}{%
  \fancyhf{}%
  \rhead{\includegraphics[width=2cm]{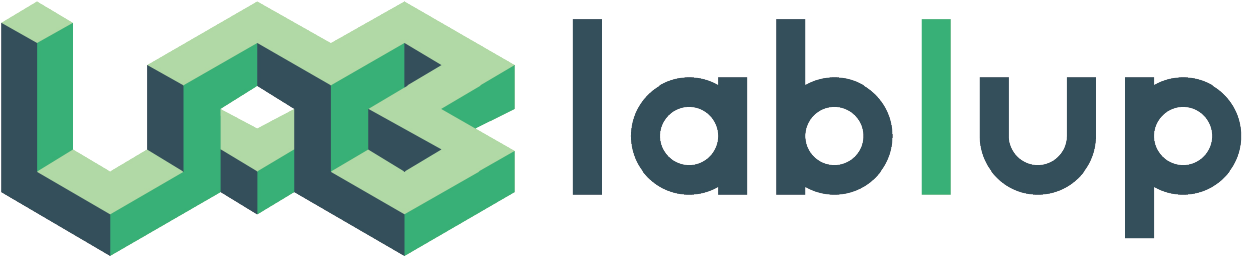}}%
  \cfoot{\thepage}%
}

\titleformat{\section}{\Large\bfseries}{\thesection}{1em}{}
\titleformat{\subsection}{\large\bfseries}{\thesubsection}{1em}{}
\titleformat{\subsubsection}{\normalsize\bfseries}{\thesubsubsection}{1em}{}

\title{\textbf{From Detection to Recovery}\\[0.3em]
{\Large Operational Analysis on LLM Pre-training\\with 504 GPUs}\\[0.5em]
\large Lablup Technical Report}
\author{\textbf{Lablup Inc.}\thanks{Please cite this work as ``Lablup Inc.~(2026)''. Full author list appears in Section~\ref{sec:authors}. For inquiries, contact us at \url{https://www.lablup.com/contact}.}}
\date{}

\begin{document}

\maketitle

\begin{abstract}
Large-scale AI training is now fundamentally a distributed systems problem, and hardware failures have become routine operating conditions rather than rare exceptions~\cite{grattafiori2024llama3, kokolis2025revisiting}. Publicly available operational evidence from production training clusters, however, remains limited. This technical report presents an empirical analysis of a 63-node NVIDIA B200 production cluster (504 GPUs), using 55 days of Prometheus time-series data and 73 days of operational logs covering 224 multi-node training sessions.

The report is based on a cross-organizational operating environment in which five parties (SKT, Upstage, Lablup, NVIDIA Korea, and VAST Data) share a unified monitoring pipeline. We document how this arrangement enabled the joint diagnosis of a 60-node-scale storage I/O bottleneck that did not appear in 2--4-node tests, illustrating a production-scale phenomenon that no single team could have isolated independently.

Using metrics collected during a months-long pre-training campaign, we perform three quantitative analyses that together yield four findings. First, for failure precursor detection, statistical analysis over 751 Prometheus metrics and 10 XID-identified GPU failures shows that no single metric is consistently dominant across failure types, which motivates a multi-signal detection strategy. Second, for checkpoint I/O profiling, analysis of 523 checkpoint events traces the save/load path from GPU VRAM to the NFS server. Restart loading reaches an average NFS read throughput of 21.5\% of the storage's maximum read bandwidth (700\,GB/s), while checkpoint-save bursts reach an average write throughput of 16.0\% of the storage's maximum write bandwidth (250\,GB/s); NFS/RPC request queueing and transport-layer backlog rise together along this path. Third, node-exclusion pattern analysis over 224 multi-node training sessions across 73 days shows a concentrated distribution in which the top 3 of 63 nodes account for more than 50\% of all exclusions. Fourth, auto-retry chain analysis quantifies a 33.3\% chain success rate over 12 chains (73 attempts total), 2.7$\times$ higher than the 12.5\% rate for manual recovery, with a median automatic retry interval of 11 minutes (IQR 10--11 min).

All analyses are grounded in production infrastructure that provides workload management at the session level, GPU-centric scheduling, and unified observability.

\end{abstract}

\newpage
\tableofcontents
\newpage

\section{Introduction}

\subsection{Background}

Once model sizes exceed 100 billion parameters, training becomes a long-running distributed systems exercise rather than an isolated algorithmic workload. A single run can require hundreds of GPUs operating in lockstep for weeks, which invalidates the classical assumption of stable underlying infrastructure. The Solar Open technical report~\cite{upstage2026solar}, for example, documents the following conditions during training of a 102-billion-parameter MoE model on a 60-node NVIDIA B200 cluster:
\begin{itemize}
    \item Persistent compatibility issues during early B200 deployment, including graph compilation failures caused by the absence of CUDA 13.0 support in Triton and ScaledDotProductAttention backend errors~(\cite{upstage2026solar}, Section~3.3.4)
    \item Performance degradation during multi-node scaling with FSDP2: TPS dropped from 5,500 to 4,267 when scaling from 16 to 60 nodes, requiring iterative tuning via HSDP to recover throughput~(\cite{upstage2026solar}, Section~3.3.2)
    \item Performance instability caused by router dtype mismatch after sigmoid operations (13.7\% speedup upon fix), unnecessary group GEMM padding overhead (14.5\% performance improvement with fast-path bypass), and gradient norm instability due to excessive token batching~(\cite{upstage2026solar}, Sections~3.3.3--3.3.4)
    \item Data loading bottleneck where I/O lock contention caused initialization to take over 8 hours~(\cite{upstage2026solar}, Section~3.3.5)
\end{itemize}
These examples show that large-scale training must be managed as a systems problem in which interruptions, restarts, and performance variability are expected. Infrastructure and orchestration therefore deserve the same analytical attention as model design.

\subsection{Problem Definition}

Large-scale AI infrastructure faces three tightly coupled challenges.

\paragraph{Low resource utilization.} Despite GPUs being expensive and scarce resources, actual utilization often remains at low levels due to static allocation policies and conservative operational practices. An analysis of Microsoft's production GPU clusters reported median GPU utilization of approximately 52\%~\cite{jeon2019analysis}.

\paragraph{Scalability--stability conflict.} As cluster scale increases, hardware failures, network latency issues, and driver errors become more frequent, raising the likelihood that training jobs are completely interrupted. Meta reported that hundreds of unexpected interruptions occurred during the 16K-GPU Llama~3 pre-training, with the majority attributable to hardware issues~\cite{grattafiori2024llama3} (we discuss this in detail in Section~\ref{sec:failure-modes}).

\paragraph{Operational complexity.} The diversity of framework, driver, and library combinations undermines environment reproducibility and introduces variability in experiment quality. MegaScale's deployment experience confirms that managing such software heterogeneity at scale constitutes a persistent operational burden~\cite{jiang2024megascale}.

These challenges reinforce one another and therefore require an integrated infrastructure-level response.

\subsection{Contributions}

This report analyzes data collected from August through December 2025 during Solar Open training on a production NVIDIA B200 cluster (63 nodes, 504 GPUs). Our contributions fall into two groups: one research-setting contribution that establishes the operational environment for the study, and four quantitative findings derived from production data.

\paragraph{Research setting.}
\begin{enumerate}[label=(S\arabic*),leftmargin=*]
    \item \textbf{Cross-organizational operating environment.} We document the five-organization collaborative environment (SKT, Upstage, Lablup, NVIDIA Korea, VAST Data) and a 60-node-scale storage I/O bottleneck that emerged only at production scale, illustrating that large-scale training phenomena cannot be predicted from 2--4-node pre-tests and that single-team monitoring is structurally insufficient for root-cause identification at this scale (Section~\ref{sec:cross-org-setting}).
\end{enumerate}

\paragraph{Quantitative findings.}
\begin{enumerate}[label=(F\arabic*),leftmargin=*]
    \item \textbf{Failure precursor detection.} For 10 XID-identified GPU failures, we apply statistical analysis to 751 Prometheus metrics. We confirm that no single metric is consistently distinctive across failure types and report ongoing time-series ML modeling to improve the pre-XID detection rate (Section~\ref{sec:precursor-analysis}).

    \item \textbf{Full-stack profiling of checkpoint I/O.} From 55 days of operational data, we provide a quantitative analysis of 523 checkpoint events and profile the save/load path from GPU VRAM to the NFS server using Prometheus metrics. Across 20 restart-loading events and 406 checkpoint-save bursts, the measured throughput averages 21.5\% of the storage's maximum read bandwidth and 16.0\% of the write bandwidth, with concurrent increases in NFS/RPC request queueing and transport-layer backlog (Section~\ref{sec:checkpoint-io}).

    \item \textbf{Analysis of node exclusion patterns.} Analyzing node exclusion frequencies across 224 multi-node training sessions over 73 days, we identify a concentrated distribution---the top 3 of 63 nodes (gpu074, gpu119, gpu086) account for more than 50\% of all exclusions---and discuss its operational implications (Section~\ref{sec:node-exclusion}).

    \item \textbf{In-depth evaluation of automated failure recovery.} Analyzing 12 auto-retry chains (73 attempts in total), we quantify chain success rate (33.3\%, 2.7$\times$ higher than manual recovery), retry-interval predictability (median 11 min, IQR 10--11 min), and downtime reduction (median 1.9 h vs.\ 3.3 h manual). We also analyze the limits in structural failures (Section~\ref{sec:auto-retry-chain}).
\end{enumerate}

\section{Failure Modes in Large-Scale Training}
\label{sec:failure-modes}

This section characterizes failures at cluster scale and derives the corresponding infrastructure requirements from recent production reports~\cite{upstage2026solar, grattafiori2024llama3, jiang2024megascale}.

\subsection{Frequent Failure Characteristics of Large-Scale Training}
Failures are an expected property of large-scale training environments~\cite{kokolis2025revisiting, grattafiori2024llama3}. Failure frequency rises with cluster scale. Meta reported 419 unexpected interruptions during the 54-day pre-training of the Llama~3 405B model on a cluster of up to 16{,}384 H100 GPUs~\cite{grattafiori2024llama3}. Meta's RSC-1 and RSC-2 clusters, which together comprise roughly 24{,}000 A100 GPUs, likewise exhibited failure frequencies that scaled with cluster size~\cite{kokolis2025revisiting}. Erben et al.\ further estimated that, at current GPU failure rates, a 100{,}000-GPU cluster would experience a failure roughly every 30 minutes~\cite{erben2024hardware}.

\subsection{Failure Characteristics of Large-Scale Clusters}

In large-scale distributed training, faults originating in a small number of nodes can affect the entire cluster. Because multi-node training synchronizes workers tightly, a single GPU failure or communication error can interrupt the whole job. A common operational response is \emph{node exclusion}, that is, withholding specific nodes from multi-node allocations. Node exclusion reflects a mixture of confirmed hardware failures, observed performance degradation, and preventive operator judgment, and therefore does not map one-to-one to hardware defects~\cite{kokolis2025revisiting}. Section~\ref{sec:node-exclusion} analyzes node exclusion patterns in our production cluster in detail.

To place our cluster's failures in context, we compare them with large-scale references. Table~\ref{tab:failure-taxonomy} reproduces the failure taxonomy used by ByteDance's Minder system~\cite{deng2025minder}, which classifies failures in production GPU clusters of roughly 1{,}500 nodes (more than 10{,}000 GPUs). Minder relies on system-wide metrics that include CPU and GPU utilization, PFC counters, network throughput, disk I/O, and memory.

\begin{table}[H]
\centering
\caption{Failure taxonomy of the ByteDance Minder system}
\label{tab:failure-taxonomy}
\resizebox{\linewidth}{!}{%
\begin{tabular}{@{}lllr@{}}
\toprule
\textbf{Category} & \textbf{Failure Type} & \textbf{Description} & \textbf{Ratio (\%)} \\
\midrule
\multirow{7}{*}{Intra-host HW} & ECC errors & GPU memory data corruption or loss & 38.9 \\
 & PCIe downgrading & Reduced transfer rate due to PCIe link failure & 6.6 \\
 & NIC dropout & NIC unrecognized by OS & 5.7 \\
 & GPU card dropout & GPU card detached from host & 2.0 \\
 & NVLink errors & NVLink interconnect failure between GPUs & 1.7 \\
 & AOC errors & Active optical cable errors & 0.9 \\
\cmidrule(l){2-4}
 & \multicolumn{2}{l}{\textbf{Subtotal}} & \textbf{55.8} \\
\midrule
\multirow{4}{*}{Intra-host SW} & CUDA runtime errors & CUDA program execution failure & 14.6 \\
 & GPU execution errors & Page faults, OOM, or GPU hangs & 7.7 \\
 & HDFS errors & Checkpoint I/O errors & 5.7 \\
\cmidrule(l){2-4}
 & \multicolumn{2}{l}{\textbf{Subtotal}} & \textbf{28.0} \\
\midrule
\multirow{2}{*}{Inter-host NW} & Machine unreachable & SSH or VM service failure & 6.0 \\
\cmidrule(l){2-4}
 & \multicolumn{2}{l}{\textbf{Subtotal}} & \textbf{6.0} \\
\midrule
Others & --- & --- & 10.3 \\
\bottomrule
\end{tabular}}%
\vspace{2pt}
\raggedright\footnotesize Based on Minder~\cite{deng2025minder} Table~1 and Appendix~A. Observed in ByteDance production GPU clusters.
\end{table}

While Minder classifies failures through analysis of system-wide metrics, our cluster uses \textit{XID error codes}---numeric fault identifiers reported by the GPU driver---recorded by the NVIDIA GPU kernel driver in \texttt{dmesg} as the primary means of failure detection. Each XID code corresponds to a specific failure type (e.g., XID~79 = GPU card dropout, XID~94 = ECC error, XID~145/149 = NVLink errors). The failure case analyses and node exclusion analyses presented in subsequent sections are based on these XID records.

Because the two systems differ in both monitoring approach and infrastructure configuration, their classification scopes do not fully overlap. The main differences are as follows:
\begin{itemize}
    \item \textbf{HDFS errors}: Not directly applicable, as our cluster uses NFS (Network File System) rather than HDFS (Hadoop Distributed File System). NFS-based checkpoint I/O issues, however, are analyzed separately in Section~\ref{sec:rpc-bottleneck}.
    \item \textbf{PCIe downgrading}: Does not generate XID errors and manifests only as bandwidth degradation, so it falls outside the XID-based analysis scope of this report (separate bandwidth monitoring would be required).
    \item \textbf{NIC dropout and AOC errors}: NICs (Network Interface Cards) and AOCs (Active Optical Cables) operate outside the GPU driver layer and are not reported as XIDs, so they are excluded from this analysis (dedicated network monitoring would be required).
    \item \textbf{CUDA runtime errors}: As application- or framework-level errors, they are excluded from the scope of hardware failure analysis in this report.
\end{itemize}

\noindent Accordingly, the analysis in this report focuses on failures detectable via XID (GPU card dropout, ECC (Error-Correcting Code) errors, NVLink failures); failure types outside the XID scope, which would require separate monitoring infrastructure, are not covered.

Table~\ref{tab:our-failure-distribution} classifies the 17 failure events recorded in our cluster during the 55-day observation period by mapping XID codes to Minder failure categories.

\begin{table}[H]
\centering
\small
\caption{Failure distribution of our cluster, mapped to Minder categories}
\label{tab:our-failure-distribution}
\begin{minipage}{\linewidth}
\centering
\begin{tabular}{@{}llrr@{}}
\toprule
\textbf{Failure Type (Minder Category)} & \textbf{XID Code} & \textbf{Count} & \textbf{Ratio (\%)} \\
\midrule
NVLink errors & 145, 149 & 5 & 29.4 \\
ECC errors & 94 & 2 & 11.8 \\
GPU card dropout & 79 & 2 & 11.8 \\
GPU execution errors & 119 & 1 & 5.9 \\
Machine unreachable & --- & 2 & 11.8 \\
Others (e.g., performance degradation) & --- & 5 & 29.4 \\
\midrule
\textbf{Total} & & \textbf{17} & \textbf{100.0} \\
\bottomrule
\end{tabular}
\vspace{2pt}

\raggedright\footnotesize 63-node cluster (504 B200 GPUs), 55-day observation period. Classified by mapping XID codes to Minder failure categories.
\end{minipage}
\end{table}

The two distributions differ in their dominant failure category. In Minder, ECC errors (38.9\%) constitute the most frequent category, whereas in our cluster, NVLink errors (29.4\%) are the most prevalent. XID codes explicitly record NVLink failures (XID~145/149), while ECC events have a small sample size due to the shorter observation period. The Others category (29.4\%) includes operational-level events such as performance degradation that do not directly map to XID codes. These differences arise from variations in monitoring strategy, observation period, and workload size distribution. Section~\ref{sec:case-studies} analyzes individual failure cases, and Section~\ref{sec:node-exclusion} examines how these failures translate into node exclusion patterns over the observation period.

\paragraph{Fail-slow faults and straggler detection.}
The taxonomy above emphasizes \textit{fail-stop} faults, in which a GPU or node stops functioning entirely. Our node exclusion data, however, reveals a second class of failures: nodes such as gpu074 and gpu119\footnote{In this report, ``gpuXXX'' denotes a node identifier within the cluster (e.g., gpu074 = node 74), while ``GPU\#N'' denotes an individual GPU index (0--7) within a node. Each node houses 8 B200 GPUs.} were repeatedly excluded because their training speed had degraded. These are \textit{fail-slow} faults, in which a component remains operational but slows down enough to impair the job. Because distributed training synchronizes workers every iteration, a single slow node can delay the entire run. Fail-slow faults are correspondingly harder to detect than fail-stop faults because they do not necessarily emit explicit error codes.

Recent production studies suggest that this pattern is widespread. Wu et al.~\cite{wu2024falcon} reported that, in 10{,}000+ GPU clusters, 59\% of large-scale training jobs (512--1{,}024 GPUs) experienced fail-slow stragglers and suffered an average job completion delay of 34.59\%. Lin et al.~\cite{lin2025whatif} found that 42.5\% of jobs in production LLM training clusters were affected by stragglers, wasting 10.4\% of total GPU hours; that study attributed the dominant causes to workload-level imbalances such as pipeline-stage skew and garbage collection pauses rather than to hardware faults. In our cluster, the lack of per-iteration throughput instrumentation meant that operators could identify fail-slow nodes only after noticing speed differences across sessions. Section~\ref{sec:future-work} outlines the monitoring extensions needed to replace this reactive process.

\subsection{XID Error Classification and Recovery Strategies}

NVIDIA GPUs report hardware and software errors through XID error codes~\cite{nvidia2026xiderrors}. Since different XID codes require fundamentally different responses, correctly classifying these codes is a prerequisite for failure attribution. Table~\ref{tab:xid-classification} classifies the XID codes observed in the operational cluster by NVIDIA-defined resolution types.

\begin{table}[H]
\centering
\caption{XID error code classification by resolution action}
\label{tab:xid-classification}
\begin{tabularx}{\linewidth}{@{}lXll@{}}
\toprule
\textbf{XID} & \textbf{Description} & \textbf{Resolution} & \textbf{Action} \\
\midrule
\multicolumn{4}{@{}l}{\textit{Hardware action required (node/GPU reset needed)}} \\
79  & GPU fell off the bus  & RESTART\_BM  & Node reboot \\
119 & GSP RPC timeout             & RESET\_GPU   & GPU reset \\
145 & NVLink RLW error            & RESET\_GPU   & GPU reset \\
149 & NVLink NETIR error          & RESET\_GPU   & GPU reset \\
\midrule
\multicolumn{4}{@{}l}{\textit{Application level (process restart sufficient)}} \\
31  & GPU memory page fault       & RESTART\_APP & Session restart \\
43  & GPU processing halted      & RESTART\_APP & Session restart \\
94  & Contained ECC error         & RESTART\_APP & Auto-corrected \\
\bottomrule
\end{tabularx}
\end{table}

This classification is reflected in Backend.AI's failure handling strategy (Section~\ref{sec:failure-recovery}). XID errors requiring hardware action (79, 119, 145, 149) trigger node isolation and session migration to spare nodes, while application-level errors (31, 43, 94) are handled through automatic retries without excluding the affected node. In practice, failure rates also depend on the system stack (OS kernel, GPU driver, firmware version) and workload intensity.

\section{Operational Infrastructure}
\label{sec:design-requirements}

Given the failure characteristics discussed in Section~\ref{sec:failure-modes}, the infrastructure must satisfy two core requirements:

\begin{enumerate}
    \item \textbf{Failure handling through detection, isolation, and recovery:} Detection operates across multiple layers---hardware (GPU ECC errors, temperature, power), process (container status, OOM (Out of Memory)), application (training progress, loss trajectory), and network (NVIDIA Collective Communications Library (NCCL) timeouts, bandwidth degradation). Application-level concerns such as loss divergence are delegated to the training framework. When a fault is detected, the affected node or GPU is isolated from the scheduling pool to limit the scope of its impact. During recovery, the system allocates replacement resources and restarts the session, with checkpointing delegated to the training framework, minimizing total training throughput loss.

    \item \textbf{Session-level lifecycle management:} The lifecycle of a training job is managed at the session level. A session is a logical unit that spans one or more containers across multiple nodes and is tied to training state, including checkpoints and optimizer state. When a session is restarted, the job resumes from the last checkpoint. The system reliably tracks and persists state transitions to support accurate recovery and auditing.
\end{enumerate}

Both requirements assume that GPUs, rather than CPUs, are the primary scheduling resource and that CPU and memory allocations are derived from GPU placement. Traditional CPU-centric orchestration does not make this assumption.

This section describes the infrastructure components that implement those requirements. The infrastructure layer sits below the training frameworks (PyTorch, DeepSpeed, Megatron-LM, and others) and provides environment provisioning, resource allocation, checkpoint storage, and failure alerting. To ground the operational analyses in Section~\ref{sec:case-studies}, we first summarize the cluster hardware (Section~\ref{sec:production-cluster}), then describe the session abstraction and recovery mechanism (Section~\ref{sec:session-recovery}), and finally explain the multi-layer monitoring pipeline used for precursor analysis (Section~\ref{sec:monitoring-observability}). We focus on the components referenced directly in the analysis sections; further implementation details are deferred to Appendix~\ref{sec:appendix-architecture}.

\subsection{Production Cluster}
\label{sec:production-cluster}

The production cluster used throughout this report is a 63-node NVIDIA HGX B200 system. Table~\ref{tab:hardware-config} summarizes the hardware configuration of the cluster.

\begin{table}[H]
\centering
\caption{Production cluster hardware configuration}
\label{tab:hardware-config}
\begin{tabular}{@{}lp{0.65\textwidth}@{}}
\toprule
\textbf{Component} & \textbf{Specification} \\
\midrule
Nodes & 63 (HGX B200): 60 for training, 3 spare \\
GPU & 8$\times$ NVIDIA B200 per node (504 total) \\
GPU Memory & 192\,GB HBM3e per GPU \\
System RAM & 2\,TiB DDR5 per node \\
Intra-node GPU Interconnect & 5th-gen NVLink, 1.8\,TB/s bidirectional \\
Inter-node Interconnect & 8-port $\times$ 400G InfiniBand (3.2\,Tbps/node) \\
Storage Network & 200G RoCE (RDMA over Converged Ethernet) \\
Storage & VAST Data E-Box, 2\,PiB, NFS mount \\
Storage Maximum Bandwidths & Cluster-aggregate read 700\,GB/s, write 250\,GB/s \\
\bottomrule
\end{tabular}
\end{table}

Model parameters, activations, and gradients reside in 192\,GB of HBM3e per GPU~\cite{nvidia2024blackwellbrief}.

During each training iteration, data traverses the following path. In the forward and backward passes, each GPU's SMs and Tensor Cores perform computation, while the 8 GPUs within a node communicate via NVLink for tensor parallelism. Gradient synchronization (AllReduce) is carried out across all 63 nodes over InfiniBand NDR (8 ports $\times$ 400G per node); if even a single node is slow, the entire training job stalls---the straggler problem~\cite{nvidia2025superpodb200}. Checkpoint saves and data loading access the VAST Data NFS storage through a dedicated 200G RoCE network. These three traffic types---compute (InfiniBand), storage (RoCE), and management (Ethernet)---are physically separated to prevent mutual interference.

This communication structure forms the physical foundation for the subsequent analyses. The NFS/RPC request-queue analysis in Section~\ref{sec:rpc-bottleneck} addresses bottlenecks on the storage plane, while the precursor analysis in Section~\ref{sec:precursor-analysis} treats InfiniBand port counters, TCP/socket metrics, and GPU telemetry (DCGM) as signals from different network planes.

The software stack running on this hardware is summarized in Table~\ref{tab:software-stack}. The upper section lists the base container image and core libraries (CUDA, cuDNN, NCCL, PyTorch), while the lower section shows the training configuration used by the Solar Open project~\cite{upstage2026solar} on this cluster---parallelization strategy, batch size, sequence length progression, and precision format.

\begin{table}[H]
\centering
\caption{Production training software stack}
\label{tab:software-stack}
\begin{tabularx}{\linewidth}{@{}lX@{}}
\toprule
\textbf{Component} & \textbf{Version} \\
\midrule
Base Image & NGC PyTorch 25.08 \\
CUDA & 13.0.0 \\
cuDNN & 9.12.0 \\
NCCL & 2.27.7 \\
PyTorch & 2.9.0.dev20250830+cu130 \\
TorchTitan & (included in PyTorch nightly) \\
Transformers & 4.55.2 \\
Flash Attention & 2.7.4.post1 \\
Python & 3.12 \\
\midrule
\multicolumn{2}{@{}l}{\emph{Training Configuration (Solar Open~\cite{upstage2026solar})}} \\
\midrule
Parallelism & HSDP (10-node sharding group $\times$ 6 replicas) \\
Global Batch Size & 13{,}440 (28 per GPU) \\
Sequence Length & 4K $\rightarrow$ 32K $\rightarrow$ 100K (progressive scaling) \\
Precision & FP8 + bfloat16 mixed \\
\bottomrule
\end{tabularx}
\end{table}

The following subsections describe the orchestration layer's session management and node isolation mechanisms used on this cluster.

\subsection{Session Abstraction}
\label{sec:session-recovery}

Deep learning training workloads are stateful tasks that preserve optimizer parameters and learning rate schedules across iterations~\cite{grattafiori2024llama3, jiang2024megascale}. Upon failure, the container is destroyed, but the training session retains all progress up to the last checkpoint. Backend.AI reflects this distinction by using \emph{sessions}---which bundle storage volumes and lifecycle state---as the core management unit instead of containers (Table~\ref{tab:session-vs-container}).

\begin{table}[H]
\centering
\small
\caption{Comparison of container and session characteristics}
\label{tab:session-vs-container}
\begin{tabular}{@{}lll@{}}
\toprule
\textbf{Property} & \textbf{Container} & \textbf{Session (DL Training)} \\
\midrule
Lifecycle & Process exit = completion & Progress to checkpoint = completion \\
State & Stateless by design & Stateful (optimizer, gradients) \\
Restart semantics & Start from scratch & Resume from checkpoint \\
Failure impact & Process lost & Progress since last checkpoint lost \\
\bottomrule
\end{tabular}
\end{table}

This distinction is directly relevant to the automated recovery analysis in Section~\ref{sec:failure-recovery}. Since ``restart'' means checkpoint resumption rather than starting from scratch, recovery time is determined by checkpoint loading time (median 31 minutes, Section~\ref{sec:restart-loading}).

\subsection{Sokovan Scheduler}
\label{sec:sokovan}

\begin{figure}[H]
  \centering
  \includegraphics[width=0.8\linewidth]{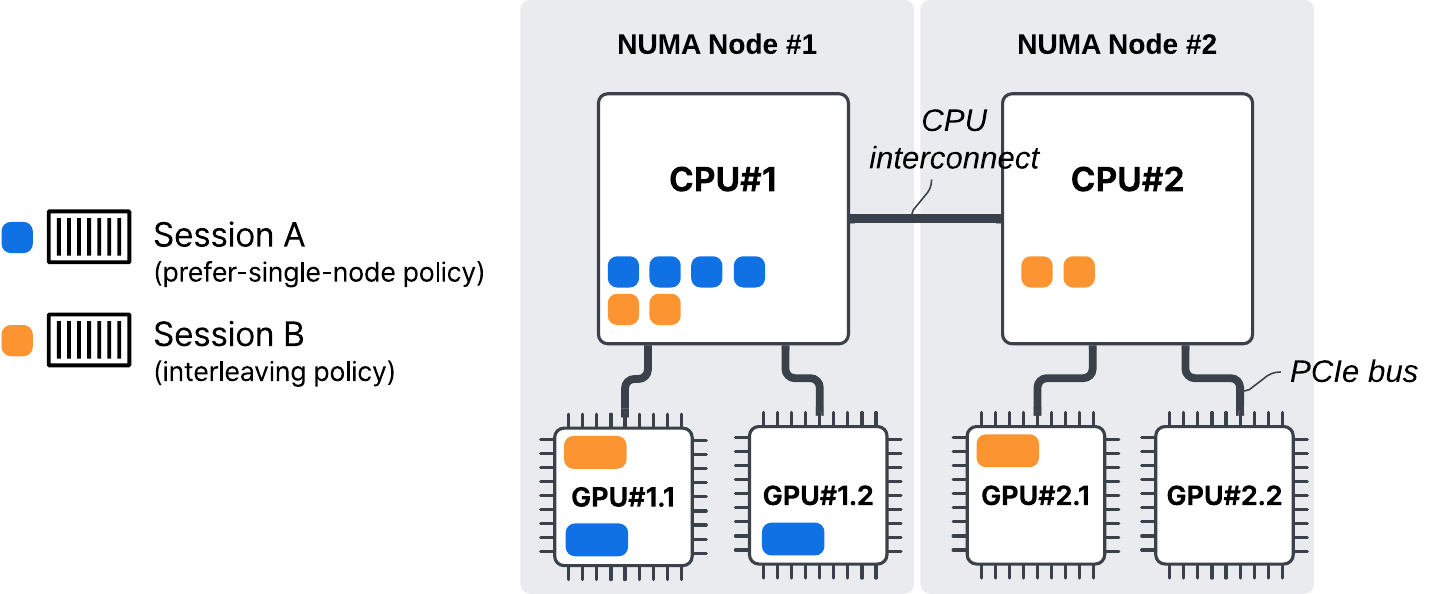}
  \caption{NUMA-aware resource allocation within a GPU server node (from~\cite{shin2023sokovan}).
  Session~A uses a \textit{prefer-single-node} policy, allocating all resources from NUMA Node~\#1.
  Session~B uses an \textit{interleaving} policy, spreading resources across both NUMA nodes.}
  \label{fig:numa-topology}
\end{figure}

GPU allocation for sessions is handled by the Sokovan scheduler. Scheduling operates at two levels: at the cluster level, pending sessions are evaluated against resource groups to control density and priority; at the node level, NUMA-aware placement policies allocate GPUs, CPU cores, and memory from the same NUMA node (Figure~\ref{fig:numa-topology}). Co-locating resources on the same NUMA node avoids cross-NUMA-node memory access, improving throughput by up to 1.30$\times$~\cite{amaral2017topology}.

Particularly critical for distributed training is gang scheduling. A 60-node training job must allocate all participating nodes simultaneously; partial allocation causes deadlocks during NCCL initialization. Sokovan allocates all $N$ slots or enqueues the entire request (all-or-nothing), preventing resource fragmentation where partially allocated jobs hold GPUs idle while waiting for the remaining slots. This constraint is directly related to the structural cause of auto-retry failures analyzed in Section~\ref{sec:auto-retry-limits}---repeated failure when fewer than 60 nodes are available.

\subsection{Multi-Layer Monitoring}
\label{sec:monitoring-observability}

GPU-only monitoring (DCGM) alone is insufficient to capture failure precursors. Table~\ref{tab:monitoring-comparison} contrasts DCGM-only monitoring with integrated multi-layer monitoring.

\begin{table}[H]
\centering
\small
\caption{Comparison of GPU-only monitoring and multi-layer monitoring}
\label{tab:monitoring-comparison}
\resizebox{\linewidth}{!}{%
\begin{tabular}{@{}lll@{}}
\toprule
& \textbf{DCGM Only} & \textbf{Integrated Monitoring (Scheduler + OS + DCGM)} \\
\midrule
Detection timing & After failure occurs (reactive) & Before failure occurs (precursor detection possible) \\
Observability scope & GPU chip state & Full stack including OS, network, and scheduler \\
\bottomrule
\end{tabular}}%
\end{table}

\noindent NVIDIA DCGM provides chip-level telemetry such as GPU temperature, power, and ECC errors, but XID error codes are recorded only after the GPU has already halted. In contrast, system-level metrics (TCP socket allocation, kernel memory, interrupts) and scheduler-level metrics (async task count, RPC latency) provide pathways through which anomalies in the GPU driver or NCCL communication layer surface before they appear in GPU telemetry.

The production cluster runs four Prometheus-compatible exporters per node, scraping metrics at 30-second intervals (Table~\ref{tab:exporters}).

\begin{table}[H]
\centering
\caption{Prometheus exporters deployed per node}
\label{tab:exporters}
\resizebox{\linewidth}{!}{%
\begin{tabular}{@{}lll@{}}
\toprule
\textbf{Exporter} & \textbf{Scope} & \textbf{Key Metrics} \\
\midrule
DCGM (Data Center GPU Manager)-exporter~\cite{nvidia2026dcgmexporter} & GPU & Utilization, temperature, power, ECC errors, NVLink, XID \\
node\_exporter~\cite{prometheus2026nodeexporter} & OS / Hardware & CPU, memory, disk I/O, network, InfiniBand, NFS \\
\texttt{all-smi}~\cite{lablup2025allsmi} & Cross-platform GPU & GPU power, system memory, chassis-level telemetry \\
Backend.AI & Scheduler & API latency, active sessions, RPC metrics \\
\bottomrule
\end{tabular}}
\end{table}

\noindent Across 63 nodes, approximately 751 unique metric names are collected in total. Of these, approximately 305 active metrics were used for failure analysis; metrics unrelated to the analysis scope (ZFS statistics, Go runtime internals, etc.) were excluded. Continuous collection over 55 days produced approximately 126\,GB of uncompressed raw telemetry, stored in VictoriaMetrics, a Prometheus-compatible time-series database.

\subsection{Cross-Organizational Operational Setting}
\label{sec:cross-org-setting}
\label{sec:storage-debugging}

The cluster is jointly operated by five organizations: SK Telecom (cloud operations), Upstage (model development), Lablup (Backend.AI infrastructure), NVIDIA (hardware), and VAST Data (storage). Operational metrics collected by each organization are aggregated through the unified monitoring pipeline (Section~\ref{sec:monitoring-observability}), allowing application, scheduler, network, and storage-layer states to be queried and analyzed on a common timeline. This study analyzes operational cases in addition to this unified observability substrate, because some causes became visible only when members compared signals across layers that were not evident from any single metric stream alone.

\paragraph{An illustrative case: storage I/O bottleneck at operational scale.} The operating environment was refined during deployment of the 60-node B200 training configuration. No major issue appeared during initial validation, but an unexpected I/O bottleneck emerged after scaling to operational size. Training initialization from VAST storage, which should have completed within minutes, instead took more than 8 hours while sustaining throughput far below the theoretical limit. As also reported in the Solar Open technical report, joint validation identified I/O lock contention as a primary cause~\cite{upstage2026solar}.

At first, the cause could not be explained from any single layer alone. Application logs showed only delayed training initialization, while storage metrics by themselves did not reveal which access pattern was responsible. After correlating application, scheduler, network, and storage metrics on the same timeline, the bottleneck was traced to the gap between the large sequential I/O intended by the application and the fragmented small random I/O that actually reached the storage layer.

Joint diagnosis across the model-development, infrastructure, hardware, and storage teams traced the problem to a mismatch between the expected large sequential I/O pattern and the fragmented small random I/O pattern that actually reached the storage layer, saturating the distributed metadata service. Each node's storage NIC receive rate (approximately 4--10\,GiB/s) looked unremarkable in isolation, but the aggregate access pattern generated simultaneously by 60 nodes overwhelmed the metadata path. Application-side file sharding (Arrow files partitioned by rank), combined with storage-side changes such as asynchronous deletion in place of synchronous \texttt{rm}/\texttt{unlink} and readahead tuning, reduced initialization time from more than 8 hours to less than 8 minutes.

\paragraph{Implications for the analyses that follow.} Two points from this case set the methodological baseline for Section~\ref{sec:case-studies}. First, performance characteristics observed at a 2--4-node scale do not predict behavior at 60 nodes; storage and metadata bottlenecks of this kind manifest only at production scale, so small-scale pre-tests are structurally insufficient for identifying them. Second, isolated monitoring by any single team is inadequate for root-cause identification at this scale; the shared metrics pipeline across organizations is what makes the systematic operational analysis in Section~\ref{sec:case-studies} tractable. The quantitative analyses that follow rest on both conditions.

\section{Operational Data Analysis}
\label{sec:case-studies}

This section analyzes four operational cases based on data collected from the production environment: cross-organizational debugging, failure precursor analysis, checkpoint/NFS analysis, and automated recovery analysis. Of the overall training campaign, the 55-day interval preserved as Prometheus time-series data is the basis for the quantitative analyses. Section~\ref{sec:precursor-analysis} analyzes metric behavior at failure time, Section~\ref{sec:checkpoint-io} analyzes NFS storage I/O and the full-stack checkpoint data path, and Section~\ref{sec:failure-recovery} quantitatively evaluates node exclusion patterns and automated recovery mechanisms.

The purpose of the analysis is to construct an operational flow from detection to diagnosis to recovery, and to show how monitoring signals connect to the actual recovery pipeline. Section~\ref{sec:cross-org-setting} has already discussed why these phenomena became visible only at operational scale through the cross-organizational debugging case.

\subsection{Failure Detection and Precursor Analysis}
\label{sec:precursor-analysis}

The previous section described a case in which the root cause was identified only after a failure became visible. This section asks whether anomalies can instead be detected before the failure manifests.

Table~\ref{tab:our-failure-distribution} reports failures at the event level. For precursor analysis, we refine those records to the node$\times$time-point level and identify 21 failures across 14 cluster downtimes\footnote{Multiple nodes may experience failures simultaneously during a single downtime. For example, during the 10/23 downtime, gpu085 and gpu122 each failed independently.}. These cases comprise 13 GPU hardware failures (10 with XID detection and 3 without), 4 fail-slow events, and 4 failures of unknown cause. We analyze the 10 cases for which XID errors immediately identified both the faulty node and the failure time (Section~\ref{sec:precursor-methodology}); in the remaining 11 cases, the absence of XID records made automatic localization difficult.

\subsubsection{Analysis Scope and Failure Classification}
\label{sec:precursor-methodology}

Precursor analysis requires that the faulty node and time point be identified. Of the 21 failures identified above, the 10 cases with node and time point identified by XID errors were selected for analysis (Table~\ref{tab:precursor-overview}). Failure types were classified based on XID error codes; when multiple XIDs occurred simultaneously in a single case, it was included in all applicable types (e.g., gpu071 falls under both NVLink and Bus Fault). All 10 cases were GPU hardware-related failures, with NVLink errors (XID 145/149) being the most frequent at 6 cases.

\begin{table}[htbp]
\centering
\caption{Overview of 10 GPU failure cases (out of 21) for which the faulty node and time point were identified by XID errors. Failure types are classified based on XID error codes. Session elapsed time indicates the time from training session start to failure occurrence.}
\label{tab:precursor-overview}
\resizebox{\linewidth}{!}{%
\begin{tabular}{@{}lllll@{}}
\toprule
\textbf{Node} & \textbf{Date} & \textbf{Failure Type} & \textbf{XID} & \textbf{Session Elapsed Time} \\
\midrule
gpu071 & 10/20 & NVLink + Bus Fault & 79, 145 & 157.6h \\
gpu085 & 10/23 & NVLink Error & 145, 149 & 0.8h \\
gpu122 & 10/23 & ECC Error & 94 & 4.7h \\
gpu085\textsuperscript{$\dagger$} & 10/23 & NVLink Error & 145, 149 & 0.7h \\
gpu116 & 10/25 & Bus Fault + ECC & 79, 94 & 37.8h \\
gpu071 & 11/9 & NVLink + Bus Fault & 79, 145 & 44.8h \\
gpu096 & 11/17 & ECC Error & 94 & 165.3h \\
gpu123 & 11/20 & NVLink Error & 145, 149 & 1.9h \\
gpu068 & 11/24 & NVLink Error & 145, 149 & 15.3h \\
gpu071 & 11/29 & GSP RPC Timeout & 119 & 62.1h \\
\bottomrule
\multicolumn{5}{@{}l}{\textsuperscript{$\dagger$}\,gpu085 experienced failures in two separate sessions on the same day (10/23) and is included as independent cases.}
\end{tabular}%
}
\end{table}

\subsubsection{Precursor Patterns by Failure Type}
\label{sec:precursor-results}

Because 60 nodes execute the same workload concurrently, anomaly detection can be framed as deviation from the peer distribution. We therefore test whether the faulty node's metrics depart significantly from the distribution observed across the remaining 59 healthy nodes. The discussion below groups failures by XID code and presents representative metric patterns for each failure type. The main goal is to show that automated faulty-node detection is feasible when the failing node diverges clearly from its peers.

\begin{figure}[H]
  \centering
  \includegraphics[width=\linewidth]{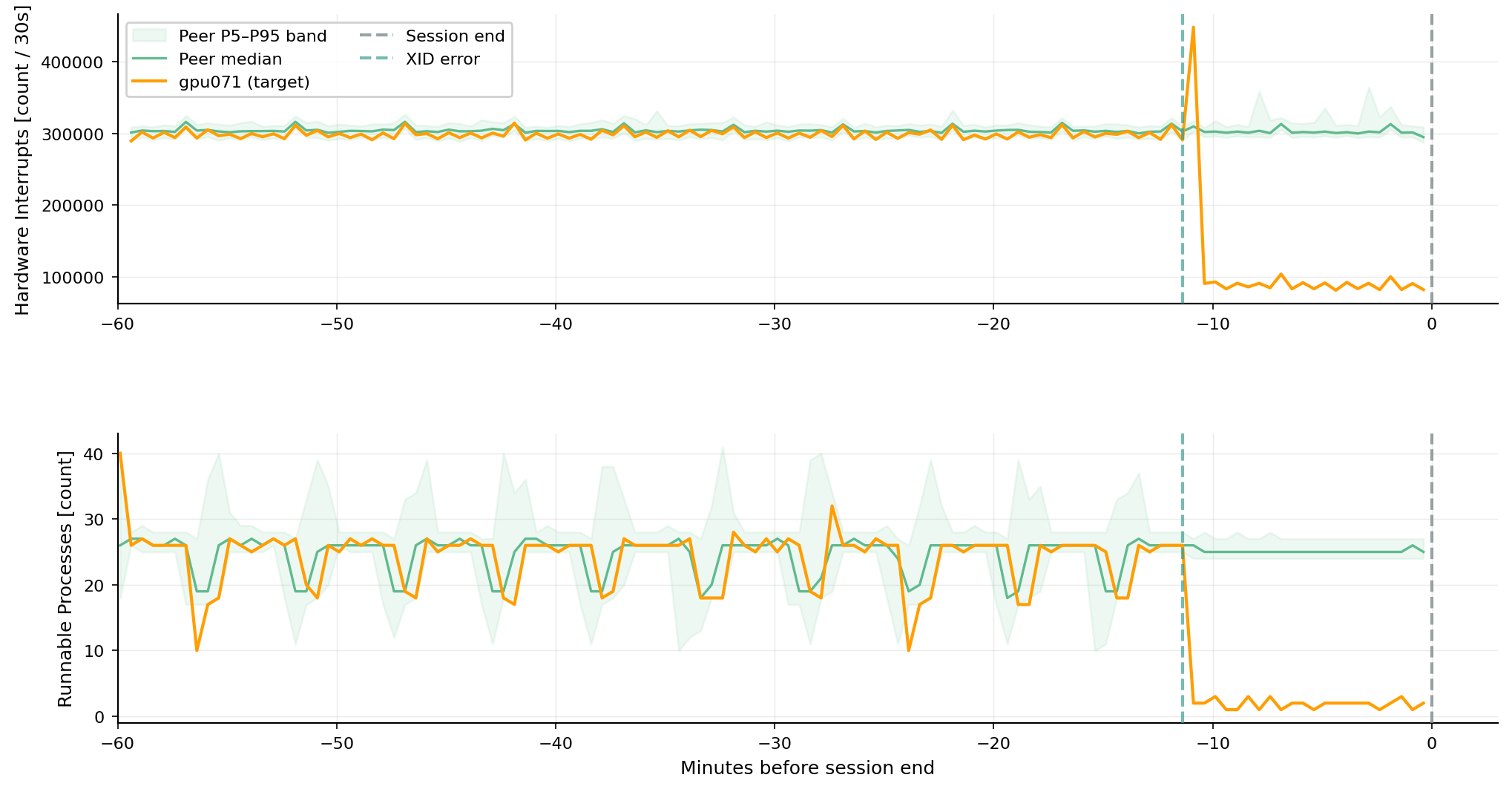}
  \caption{NVLink + Bus Fault (gpu071, 2025-10-20). Top: host CPU interrupts (\texttt{node\_intr\_total}, 30-second counter increment); drops sharply at XID time point from approximately 300K (peer) to about 70K--100K. Bottom: runnable process count (\texttt{node\_procs\_running}, instantaneous value); drops to 0 at XID time point as the worker process terminates.}
  \label{fig:precursor-se-nvlink}
\end{figure}

\paragraph{NVLink-related failures (XID 145/149, 6 cases).} Figure~\ref{fig:precursor-se-nvlink} shows the NVLink + Bus Fault case for gpu071 (10/20). The top panel is the count of interrupts handled by the host CPU (\texttt{node\_intr\_total}, 30-second counter increment), and the bottom panel is the number of currently runnable processes (\texttt{node\_procs\_running}, instantaneous value). For both metrics, peer nodes remain stable throughout training, while gpu071 deviates clearly at the XID time point. The interrupt count drops sharply at the XID time point from approximately 300K (peer) to around 70K--100K, consistent with no further interrupts being generated on the device after the GPU was disconnected from the bus due to the NVLink error. The number of runnable processes is comparable to peers during training but drops to 0 at the XID time point---the training worker process terminated as the GPU halted, eliminating the runnable processes themselves. A similar pattern was observed when the failure recurred on the same node on 11/9.

\begin{figure}[H]
  \centering
  \includegraphics[width=\linewidth]{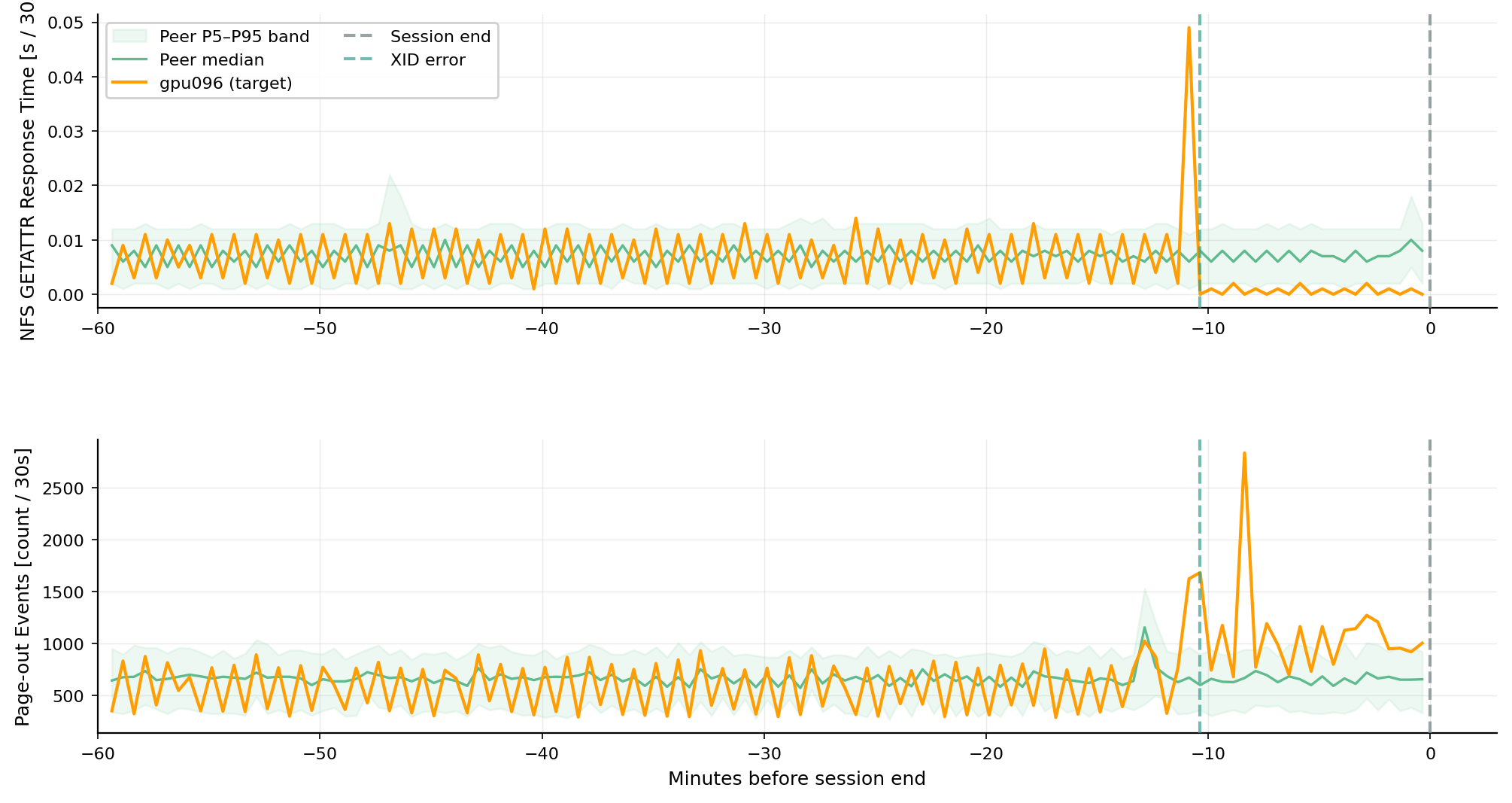}
  \caption{ECC Error (gpu096, 2025-11-17). Top: NFS GETATTR response time (\texttt{node\_mountstats\_\allowbreak nfs\_\allowbreak operations\_\allowbreak response\_time\_\allowbreak seconds\_total}, 30-second cumulative time); surges relative to peers at XID time point. Bottom: page-out events (\texttt{node\_vmstat\_pgpgout}, 30-second counter increment); surges at XID time point.}
  \label{fig:precursor-se-ecc}
\end{figure}

\paragraph{Contained Memory Error (XID 94, 3 cases).} Among ECC (Error-Correcting Code) errors, XID 94 is reported on multi-bit (uncorrectable) memory faults that ECC could not correct. Single-bit correctable errors are auto-handled by ECC and not reported as XID. Figure~\ref{fig:precursor-se-ecc} shows the ECC Error case for gpu096 (11/17). The top panel is the cumulative response time of NFS GETATTR requests (\texttt{node\_mountstats\_\allowbreak nfs\_\allowbreak operations\_\allowbreak response\_time\_\allowbreak seconds\_total}, GETATTR operation, 30-second cumulative time), and the bottom panel is the cumulative count of page-outs from memory to disk/storage (\texttt{node\_vmstat\_pgpgout}, 30-second counter increment). Both metrics show a clear surge on gpu096 relative to peers at the XID time point. We hypothesize that kernel-side cleanup work occurring just after the worker process terminated abnormally due to the ECC error---NFS revalidation triggered by file-handle reclamation and writeback flush of held dirty pages---led to the simultaneous surges in both metrics, though the exact causal mechanism cannot be determined from this data alone.

\begin{figure}[H]
  \centering
  \includegraphics[width=\linewidth]{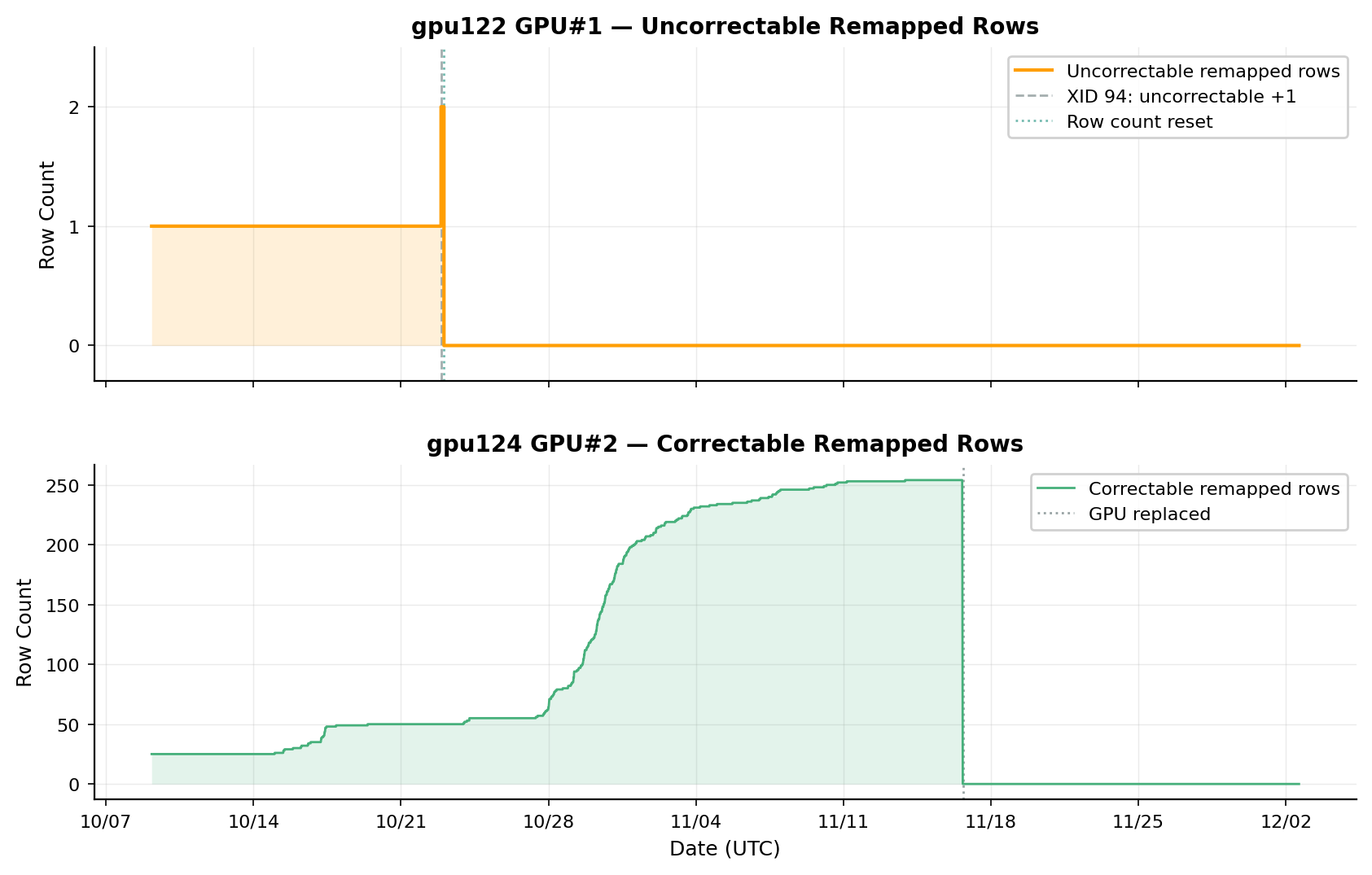}
  \caption{ECC memory row remapping timeline. Top --- gpu122 GPU\#1 uncorrectable (permanent fault) remapping: when remappings increased, XID 94 ECC error fired simultaneously, halting the GPU. Bottom --- gpu124 GPU\#2 correctable (hardware auto-corrected) remapping: accumulated up to 254 rows over 55 days at an accelerating pace, with no XID error reported. The GPU was replaced on 11/17 because it was no longer visible to the system.}
  \label{fig:remapped-rows}
\end{figure}

The memory row remapping counter provided by DCGM can be used as an indicator for tracking long-term hardware degradation. When an ECC error occurs in GPU memory, the defective row is remapped to a spare row; uncorrectable remapping indicates a permanent defect. The top panel of Figure~\ref{fig:remapped-rows} shows the uncorrectable remapping trend for gpu122 (GPU\#1). XID 94 (ECC Error) occurred simultaneously at the time points when this value increased, causing the GPU to halt. Uncorrectable remapping indicates progressing memory degradation, and when uncorrectable remappings per bank reach 8, the \texttt{ROW\_REMAP\_FAILURE} flag is triggered, necessitating GPU replacement~\cite{nvidia2025gpumemerrormgmt}.

The bottom panel of Figure~\ref{fig:remapped-rows} shows the correctable remapping for gpu124 (GPU\#2). It accumulated over approximately 55 days---increasing slowly at first, then surging from around October 28 to exceed 200 within a week before reaching 254 rows. No uncorrectable remapping or XID errors occurred during this period, yet the GPU was eventually replaced because it was no longer recognized by the host. NVIDIA advises that correctable errors can be ignored, as hardware corrects them automatically~\cite{nvidia2025gpumemerrormgmt}; however, a rapidly increasing trend may signal progressing memory degradation and warrant monitoring.

Across the 10 analyzed cases, no single precursor metric dominates consistently across failure types. Even within the same XID category (for example, NVLink 145/149) and even across recurrences on the same node (the two NVLink events on gpu071), the strongest signals differ. We therefore adopt a multi-signal strategy rather than relying on a single metric. To improve the pre-XID detection rate, follow-on ML modeling of multivariate time-series patterns and cross-metric correlation changes is in progress.

\subsection{Checkpoint Save and Recovery: Storage Bottleneck Analysis}
\label{sec:checkpoint-io}

Checkpoint behavior determines both the progress lost at failure time and the time required to resume training.

This section first characterizes the training I/O profile and checkpoint interval, then quantifies failure-induced loss and restart bottlenecks using W\&B runs. It finally traces the checkpoint path from GPU VRAM to the NFS server using Prometheus metrics, which allows us to observe the asynchronous checkpoint pipeline and quantify NFS/RPC request queueing.

\subsubsection{Training I/O Profile and Checkpoint Interval}
\label{sec:io-profile}

\begin{figure}[H]
  \centering
  \includegraphics[width=\linewidth]{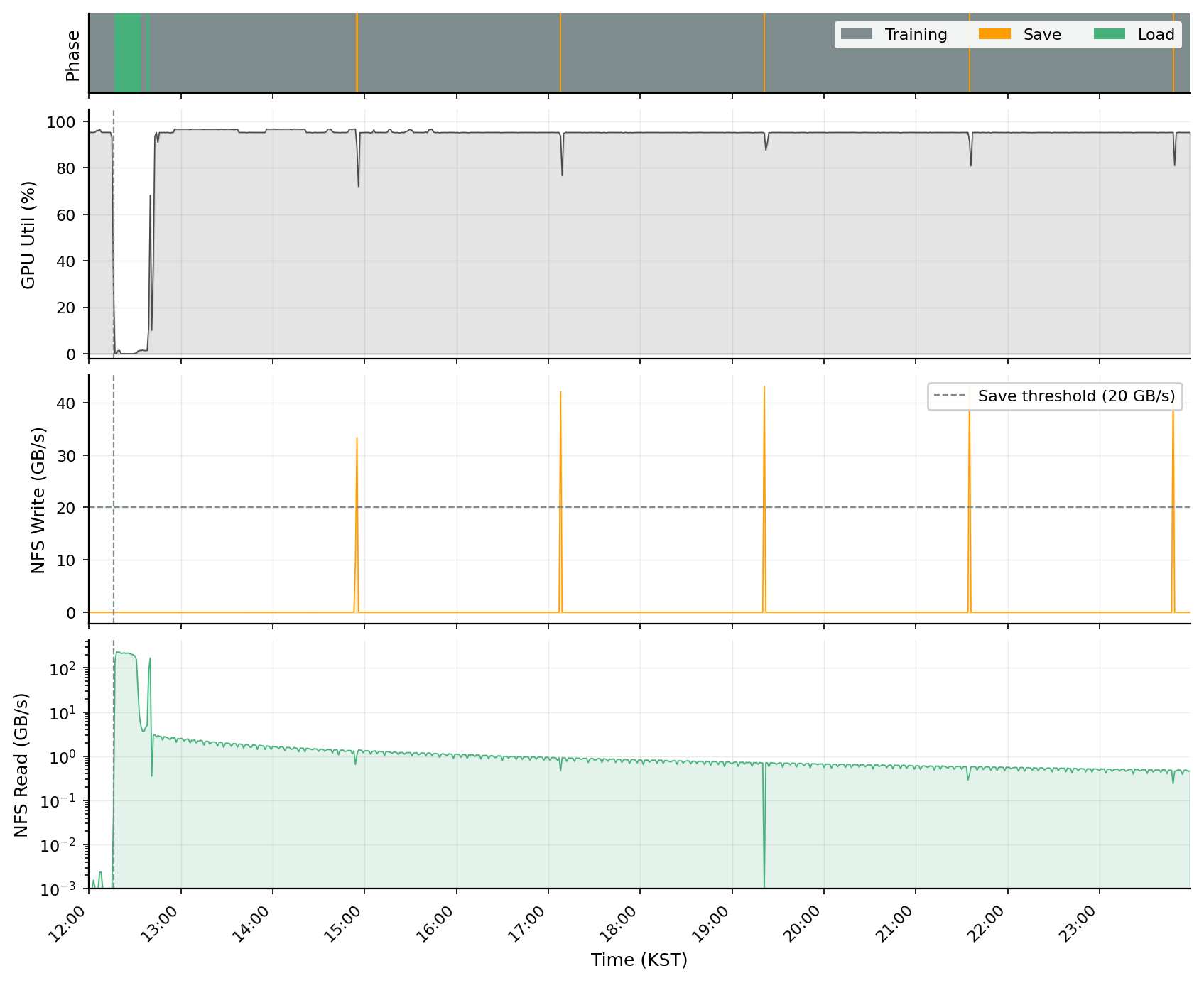}
  \caption{Training I/O profile (2025-10-25 12:00--10-26 00:00 KST). From top: training phase classification bar (gray=Training, orange=Save, green=Load), cluster-average GPU utilization (\%), cluster-aggregate NFS write rate (GB/s, gray dashed line=20\,GB/s Save detection threshold), cluster-aggregate NFS read rate (GB/s, log scale). Gray dashed line indicates session start time (12:16 KST).}
  \label{fig:training-io-profile}
\end{figure}

The time series can be partitioned into three training phases based on GPU utilization and NFS I/O patterns (Figure~\ref{fig:training-io-profile}). In decreasing order of priority, we classify intervals as Save (checkpoint save; cluster-aggregate NFS writes $> 20$\,GB/s), Load (checkpoint and data load; GPU utilization $< 50\%$ and cluster-aggregate NFS reads $> 2$\,GB/s), and Training (all remaining intervals).

During stable training, GPU utilization is maintained above 99\%, with brief dips observed at approximately 2-hour 13-minute intervals. These dips are precisely synchronized with NFS write spikes, showing that checkpoint saves temporarily pause GPU computation. The per-node write volume per checkpoint is approximately 20\,GB, remaining constant throughout the period; the cluster-aggregate peak write rate decreased from approximately 43\,GB/s early in the analysis to approximately 31\,GB/s after mid-November. The decrease in cluster-aggregate rate while per-node write volume remained constant suggests that the same amount of data was distributed over a longer duration. The precise cause (NFS server-side load, performance changes due to growing storage capacity, etc.) requires further investigation.

At session startup, NFS reads surge to approximately 230\,GB/s cluster-aggregate due to checkpoint and training data loading, with approximately 200\,GB loaded into the page cache per node over about 25 minutes. After this, NFS network traffic converges to virtually 0 during training, with all data access served from the page cache.

\begin{figure}[H]
  \centering
  \includegraphics[width=\linewidth]{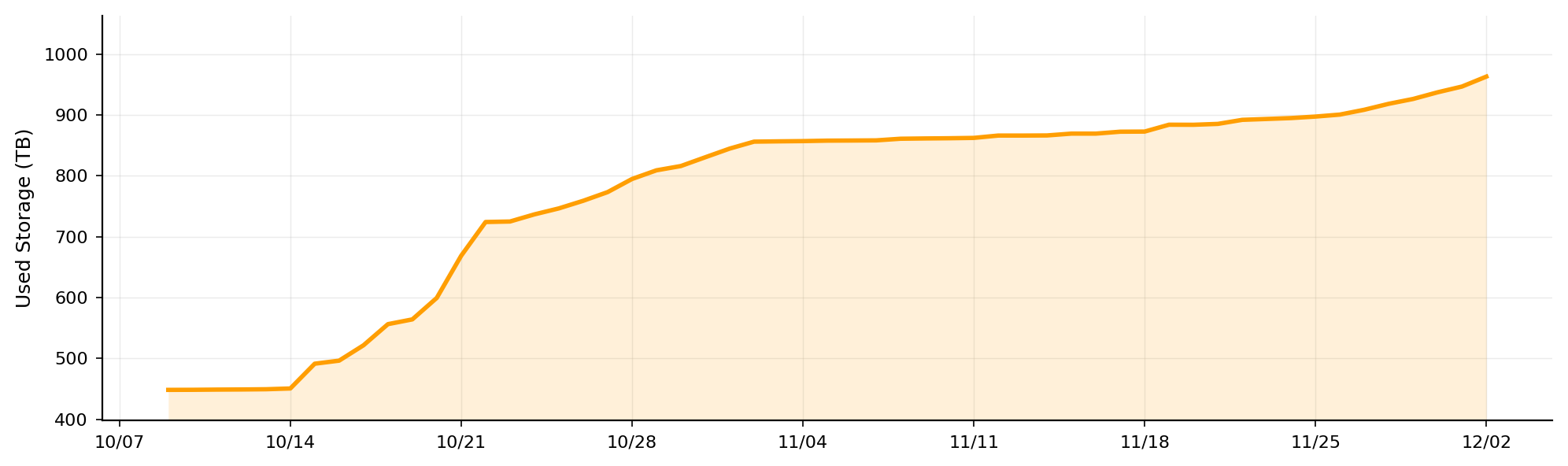}
  \caption{NFS storage usage trend over 55 days (total capacity approximately 2,252\,TB). Checkpoint file accumulation is the primary driver of growth.}
  \label{fig:nfs-storage-usage}
\end{figure}

A total of 523 checkpoint events were automatically detected over 55 days based on NFS write spikes. Checkpoints are stored on VAST Data NFS storage, and the interval varied systematically with the training configuration. According to the Solar Open technical report~\cite{upstage2026solar}(Section~3.4), training proceeds in three phases: pretraining (sequence length 4K, per-GPU batch 28, global batch 13{,}440), context extension phase 1 (32K, per-GPU batch 3, global batch 1{,}440), and context extension phase 2 (100K, per-GPU batch 1, global batch 480). The measured checkpoint intervals for each phase are as follows.
\begin{itemize}
  \item \textbf{4K sequence phase} (pretraining): median 2.23 hours (133.5 min), standard deviation 5.4 min, stable. This phase accounted for most of the analysis period (466 events).
  \item \textbf{32K sequence phase} (context extension phase 1): increased to 3.32 hours (199 min), attributed to longer step times from the extended sequence length.
  \item \textbf{100K sequence phase} (context extension phase 2): 1.36 hours (81.5 min), shorter than 32K and close to the theoretical optimum; the optimization effect is discussed below.
\end{itemize}
NFS storage usage increased from approximately 450\,TB to 963\,TB during the analysis period, an increase of approximately 510\,TB (Figure~\ref{fig:nfs-storage-usage}). Utilization changed from approximately 20\% to 43\% out of the total capacity of approximately 2,252\,TB.

\subsubsection{Failure Cost and Checkpoint Interval}
\label{sec:failure-cost}

Checkpoint interval sets a direct trade-off between save overhead and lost progress at failure time. More frequent checkpoints reduce lost work but increase save overhead, whereas longer intervals reduce overhead but increase the amount of discarded training. The Young/Daly model~\cite{young1974first, daly2006higher} provides the standard reference point for this trade-off.

This model assumes that failures occur uniformly (memorylessly) over time. In practice, however, the lost-time distribution is not uniform: operator-initiated terminations shortly after checkpoints (loss 0.05--0.1 hours) coexist with unexpected failures in the mid-to-late interval (loss 2--3 hours)~\cite{herault2024survey}, so the theoretical optima below serve as reference points for setting operational targets. Nonetheless, an operational lesson can be drawn from this analysis: checkpoint save overhead ($\delta = 18\text{--}31.7$ seconds) is small enough that the cost of shorter intervals is low, and reducing the interval to 81.5 minutes in the 100K sequence phase brought total cost (1.82\%) close to the theoretical optimum (1.72\%).

\begin{figure}[H]
  \centering
  \includegraphics[width=\linewidth]{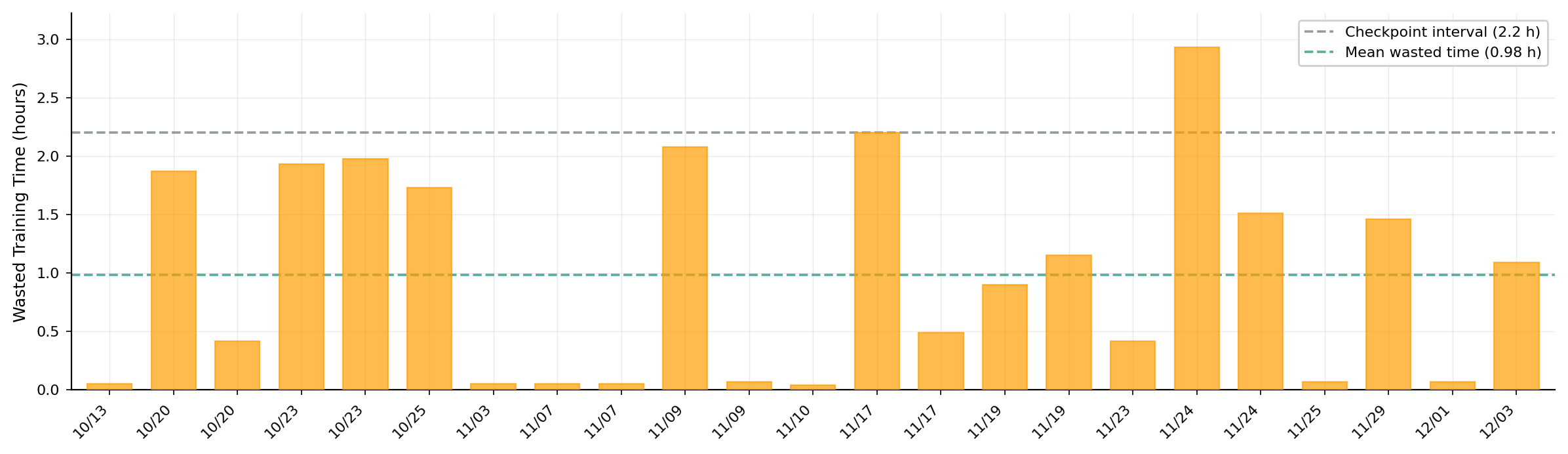}
  \caption{Lost training time per abnormally terminated W\&B run (23 cases). Gray dashed line = 4K-stage checkpoint interval (2.2 hours), teal dashed line = mean lost time (0.98 hours).}
  \label{fig:wasted-training-time}
\end{figure}

\paragraph{Measured lost time.} For 23 abnormally terminated W\&B runs\footnote{These 23 cases are abnormally terminated runs recorded in W\&B, which differ in counting unit from the 17 events in Table~\ref{tab:our-failure-distribution} (Prometheus/XID-based failure events): W\&B runs represent session terminations from the training framework perspective, while Table~\ref{tab:our-failure-distribution} counts hardware-layer XID error occurrences.}, we identified the preceding checkpoint from NFS write spike timestamps and calculated the difference from the termination time (Figure~\ref{fig:wasted-training-time}). The mean lost time was 0.98 hours, with a total of approximately 22.6 hours.

\paragraph{Interval optimization.} Checkpoint intervals involve a trade-off: shorter intervals increase save overhead ($\delta/T$), while longer intervals increase the average loss upon failure ($T/2M$). The Young/Daly model gives the optimal interval $T_{\text{opt}} = \sqrt{2\delta M}$.

Save duration $\delta$ cannot be measured directly at 30-second Prometheus sampling, so we estimated it from the number of consecutive NFS write spike samples (Table~\ref{tab:delta-estimation}). MTBF ($M$) was estimated at 56.2 hours from 1{,}294 total training hours across 24 runs divided by 23 abnormal terminations.

\begin{table}[H]
\centering
\small
\caption{Statistical estimation of checkpoint save duration ($\delta$). $\bar{N}$ is the average number of consecutive 30-second samples spanned by an NFS write spike.}
\label{tab:delta-estimation}
\begin{tabular}{@{}lrrr@{}}
\toprule
\textbf{Training Phase} & \textbf{Episodes} & $\bar{N}$ (samples) & $\delta$ \textbf{(sec)} \\
\midrule
4K sequence & 466 & 1.60 & 18.0 \\
32K sequence & 36 & 2.06 & 31.7 \\
100K sequence & 21 & 2.00 & 30.0 \\
\bottomrule
\end{tabular}
\end{table}

Table~\ref{tab:checkpoint-interval} compares costs by training phase. In the 4K and 32K phases, the actual interval was approximately 3$\times$ the theoretical optimum, with failure loss (1.98--2.95\%) dominating the cost. In the 100K phase, the interval was shortened to 81.5 minutes (1.4$\times$ the optimum), and total cost (1.82\%) was within 0.10 percentage points of the theoretical minimum (1.72\%). Save overhead remained below 0.6\% in all phases, confirming the low cost of shorter intervals.

\begin{table}[H]
\centering
\small
\caption{Checkpoint interval cost comparison by training phase ($M = 56.2$ hours).}
\label{tab:checkpoint-interval}
\begin{tabular}{@{}lrrrrr@{}}
\toprule
\textbf{Training Phase} & $\delta$ & \textbf{Actual Interval} & $T_{\text{opt}}$ & \textbf{Save Overhead} & \textbf{Total Cost} \\
\midrule
4K sequence & 18\,s & 133.5 min & 44.9 min & 0.22\% & 2.20\% \\
32K sequence & 31.7\,s & 199.0 min & 59.7 min & 0.27\% & 3.22\% \\
\textbf{100K sequence} & \textbf{30\,s} & \textbf{81.5 min} & \textbf{58.1 min} & \textbf{0.61\%} & \textbf{1.82\%} \\
\bottomrule
\end{tabular}
\end{table}

\subsubsection{Restart Loading Time and Bandwidth Utilization}
\label{sec:restart-loading}

\begin{figure}[H]
  \centering
  \includegraphics[width=\linewidth]{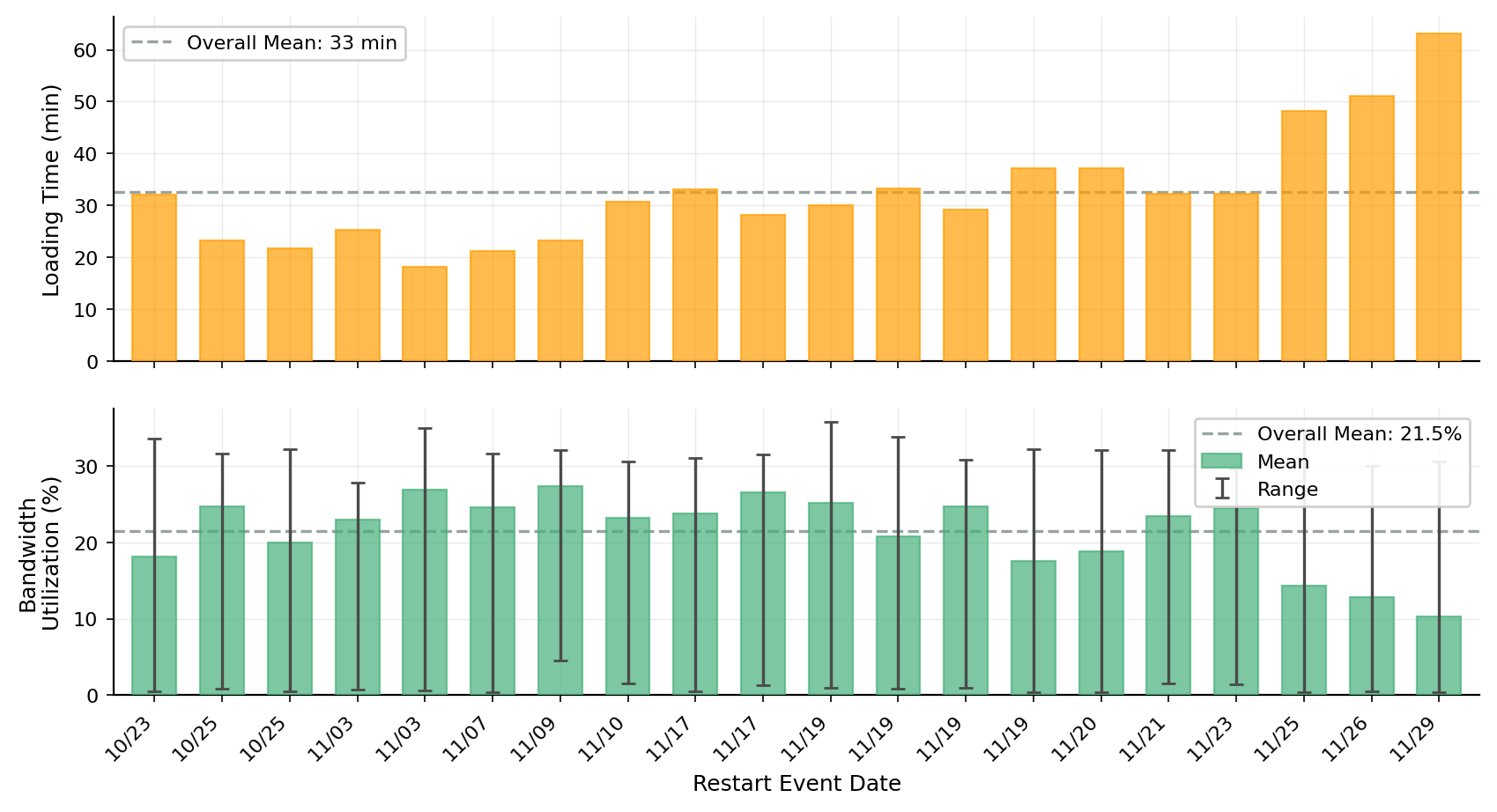}
  \caption{Session loading time and NFS bandwidth utilization (20 sessions). Top: loading time (minutes), gray dashed line = mean (33 min). Bottom: NFS read utilization relative to the storage's maximum read bandwidth (700\,GB/s). Bars = mean, error bars = min to max (based on 30-second Prometheus samples, overall mean 21.5\%), gray dashed line = mean (21.5\%).}
  \label{fig:restart-loading}
\end{figure}

Restart loading time is a first-order determinant of recovery latency (Figure~\ref{fig:restart-loading}). We measured the time required to load checkpoints and datasets at session startup across 20 sessions that lasted more than 1 hour and for which the loading phase could be identified. Loading time is defined as the interval from session start to the end of the Startup/Loading phase (Section~\ref{sec:io-profile}: GPU utilization $< 50\%$ and cluster-aggregate NFS reads $> 2$\,GB/s). The mean loading time is 33 minutes and the median is 31 minutes.

Across the 20 restart-loading events, average NFS read throughput is 150.8\,GB/s, corresponding to 21.5\% of the storage's maximum read bandwidth of 700\,GB/s. We use the storage's maximum bandwidths (700\,GB/s read and 250\,GB/s write) of the cluster configuration for the comparison and these values should not be interpreted as separately measured peak bandwidths in this study. The mean of event-level peak read throughput is 223.8\,GB/s, or 32.0\% of the same maximum bandwidths. The observed read throughput remains below the storage maximum bandwidths. A network-only upgrade from 200\,Gbps to 400\,Gbps RoCE is therefore unlikely to substantially reduce restart loading time. The following analysis examines how NFS/RPC request queueing and transport-layer backlog form during checkpoint restore.

Table~\ref{tab:nfs-summary} summarizes the key metrics from this section.

\begin{table}[H]
\centering
\small
\caption{Checkpoint I/O quantitative analysis summary (55 days, 60-node training on 63-node B200 cluster)}
\label{tab:nfs-summary}
\begin{tabular}{@{}llr@{}}
\toprule
\textbf{Category} & \textbf{Metric} & \textbf{Measured Value} \\
\midrule
\multirow{2}{*}{Storage Usage} & Start / End & 450\,TB / 963\,TB \\
 & Utilization Change & approx.\ 20\% $\rightarrow$ 43\% \\
\midrule
\multirow{5}{*}{Checkpoints} & Detected Events & 523 \\
 & Interval (4K / 32K / 100K) & 2.23 / 3.32 / 1.36\,h \\
 & Save Duration $\delta$ (4K / 32K / 100K) & 18.0 / 31.7 / 30.0\,s \\
 & Per-node Write Volume & approx.\ 20\,GB \\
 & Cluster-aggregate Peak Write (30s avg) & 31--43\,GB/s \\
\midrule
\multirow{2}{*}{Failure Cost (23 runs)} & Mean Lost Time & 0.98\,h \\
 & Total Lost Time & 22.6\,h \\
\midrule
\multirow{4}{*}{Restart Loading and I/O} & Loading Time Mean / Median & 33\,min / 31\,min \\
 & Average READ Throughput / Maximum Value & 150.8\,GB/s / 21.5\% \\
 & Mean Peak READ / Maximum Value & 223.8\,GB/s / 32.0\% \\
 & Checkpoint-save Burst Average / Maximum Value & 40.1\,GB/s / 16.0\% \\
\bottomrule
\end{tabular}
\end{table}

To understand where checkpoint I/O time is spent, we analyzed the \texttt{mmap()}, page cache, and NFS/RPC handling layers along the save and load data paths.

\subsubsection{Checkpoint Data Path: From GPU to NFS}
\label{sec:fullstack-save}

Checkpoint saving and loading follow opposite data paths. On save, the model and optimizer state in GPU VRAM is copied to a CPU-side staging buffer and then written to storage via \texttt{write()} calls, kernel writeback, and NFS WRITE RPCs. At 30-second monitoring granularity, the following signals appear in sequence: a drop in GPU utilization, increased staging buffer usage, rising dirty pages and writeback, higher NFS write traffic, and growing transport-layer backlog.

\texttt{/dev/shm} staging usage stays constant because the staging buffer is pre-allocated at training start and the same region is reused for each save. The 60 training nodes split into two groups (48 nodes at approximately 48\,GB, 12 nodes at approximately 9\,GB), and the per-node NFS write volume per checkpoint follows the same pattern (approximately 26\,GB and 2\,GB, respectively). This indicates that the larger \texttt{/dev/shm} region serves as the staging buffer in which each node holds its checkpoint shard before transmitting it to NFS. The cumulative \texttt{write()} system-call volume and the volume received by the NFS server match at 20.55\,GB per node (1,295\,GB cluster-aggregate), confirming that all saved data passed through the \texttt{write()} path.

Loading proceeds in the opposite direction. Immediately after restart, NFS READ throughput, major page faults, readahead, and transport-layer backlog rise first, after which data loaded into the page cache is copied into user buffers and GPU VRAM. In some intervals, the \texttt{read()} volume exceeds the increase in NFS reads, indicating that the page cache retained from a previous session is reused. Restart loading time is therefore not determined by network transfer time alone; page-cache population, NFS/RPC request handling, and the training framework's checkpoint restore stage all contribute.

\subsubsection{NFS/RPC Request Queueing}
\label{sec:rpc-bottleneck}

\begin{figure}[H]
  \centering
  \includegraphics[width=\linewidth]{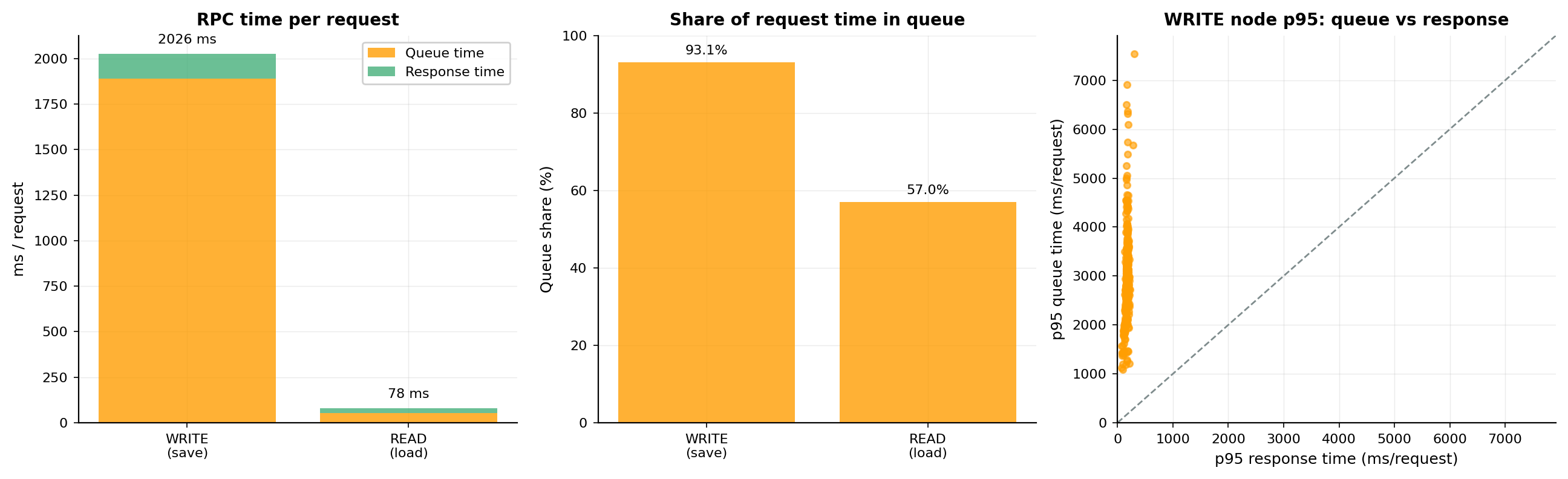}
  \caption{Comparison of NFS/RPC queue and response time. Left: average per-request queue time and response time for 406 checkpoint-save bursts and 20 restart-loading events. Center: queue-time share of total request time. Right: node-level p95 queue time versus p95 response time for checkpoint-save bursts.}
  \label{fig:rpc-bottleneck}
\end{figure}

NFS request time can be separated into queue time and response time. Queue time reflects the time a request waits in the client or transport path before transmission or processing, while response time reflects the interval after the request is sent until the response is received. We compute per-request times from Prometheus NFS operation counters (\texttt{queue\_time\_seconds\_total}, \texttt{response\_time\_seconds\_total}, \texttt{request\_time\_seconds\_total}, and \texttt{requests\_total}) across 406 checkpoint-save bursts and 20 restart-loading events (Figure~\ref{fig:rpc-bottleneck}).

\begin{table}[H]
\centering
\small
\caption{Comparison of RPC request patterns and per-request latency between save and load.}
\label{tab:rpc-comparison}
\begin{tabular}{@{}llll@{}}
\toprule
\multicolumn{2}{l}{} & \textbf{Save (WRITE)} & \textbf{Load (READ)} \\
\midrule
\multicolumn{2}{l}{Analysis unit} & 406 save bursts & 20 restart-loading events \\
\multicolumn{2}{l}{Average throughput / Maximum Value} & 40.1\,GB/s (16.0\%) & 150.8\,GB/s (21.5\%) \\
\multicolumn{2}{l}{Average RPC size} & 1{,}021\,KB/request & 440\,KB/request \\
\midrule
\multirow{3}{*}{Per-request latency} & \textbf{Total} & \textbf{2{,}026\,ms} & \textbf{78.2\,ms} \\
 & Queue time & 1{,}890\,ms (93.1\%) & 54.2\,ms (57.0\%) \\
 & Response time & 136\,ms (6.7\%) & 23.7\,ms (30.3\%) \\
\bottomrule
\end{tabular}
\end{table}

The measurements in Table~\ref{tab:rpc-comparison} show that a WRITE request takes 2.03 seconds on average. Of this, 1.89 seconds (93.1\%) is queue time, while response time is 136\,ms. The average WRITE RPC size is 1{,}021\,KB/request, close to the NFS \texttt{wsize}, and metadata operations account for only 0.85\% on average (median 0.23\%). The dominant pattern on the save path is therefore not an increase in small WRITE requests or metadata operations, but the accumulation of queue time after large WRITE RPCs are issued concurrently by many nodes.

The node-level analysis points in the same direction. Among the 406 save bursts, 87.9\% involve at least 50 nodes issuing WRITE requests concurrently, and the median write share of the top 10 nodes is 21.0\%. The reduction in average write throughput is thus a cluster-wide phenomenon. At node p95, WRITE queue time averages 3{,}020\,ms/request, whereas response time averages 166\,ms/request; p95 queue time exceeds p95 response time in every save burst.

On the load path, READ RPCs average 440\,KB/request, and 57.0\% of request time is queue time. When the loading interval is divided into early, middle, and late phases, the late-phase throughput drop appears in all 20 restart-loading events. During the late phase, READ bytes, READ requests, and open operations decrease, and queue time does not rise together with transport-layer backlog. This pattern is consistent with the end of restore being delayed by remaining checkpoint shards and residual work on a subset of ranks.

\paragraph{Bandwidth utilization and instrumentation limits.} Relative to the maximum bandwidths above, the measured 30-second average throughput is 40.1\,GB/s (16.0\%) during saves and 150.8\,GB/s (21.5\%) during loads. Save bursts complete within tens of seconds, so 30-second sampling can understate instantaneous write throughput; loading lasts for more than 20 minutes, making this effect smaller. The current data consists of client-side mountstats and Prometheus metrics. It cannot separate whether the observed queue time is formed primarily by client-side concurrency limits, the \texttt{nconnect} configuration, transport-layer backlog, NFS frontend flow control, or backend write handling inside the storage system. The observable conclusion is that NFS/RPC request queueing and transport-layer backlog are the first signals to examine on the checkpoint I/O path.

\subsection{Failure Patterns and Automated Recovery}
\label{sec:failure-recovery}

This subsection analyzes failure response in multi-node training from two complementary perspectives. Section~\ref{sec:node-exclusion} asks \emph{where} failures recur by identifying which nodes are repeatedly excluded and why. Section~\ref{sec:auto-retry-chain} asks \emph{how effectively} the system recovers once a failure occurs. The two analyses are linked: the spare-node shortage that limits auto-retry effectiveness (Section~\ref{sec:auto-retry-limits}) follows directly from the exclusion distribution.

\subsubsection{Node Exclusion Patterns}
\label{sec:node-exclusion}

Node exclusion is highly concentrated rather than uniformly distributed across the cluster. Across 224 multi-node training sessions over 73 days, the same nodes were repeatedly withheld from 60-node jobs. The cluster contains 63 GPU nodes. When a 60-node session starts, Sokovan selects from the set of nodes whose resources are currently free. Operators can deliberately exclude nodes from multi-node scheduling by pre-allocating single-node sessions to them. Because the cluster has only 3 spare nodes, and those spares are often occupied in this way, the effective node composition of large training jobs becomes nearly fixed.

\begin{figure}[H]
  \centering
  \includegraphics[width=\linewidth]{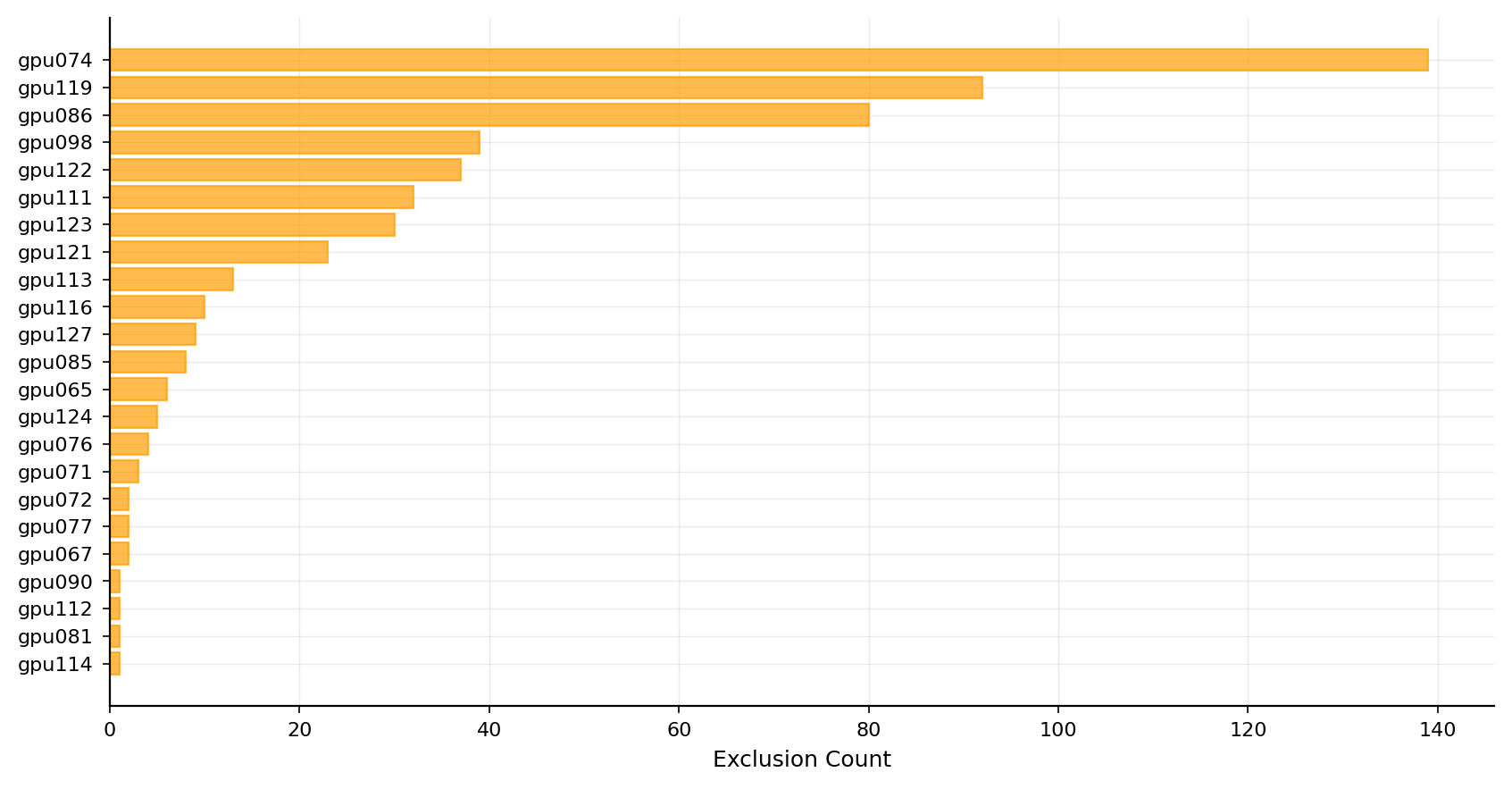}
  \caption{Node exclusion frequency across 224 multi-node training sessions over 73 days. The top 3 nodes (gpu074, gpu119, gpu086) account for over 50\% of all exclusions, showing a concentrated distribution.}
  \label{fig:node-exclusion-frequency}
\end{figure}

Figure~\ref{fig:node-exclusion-frequency} shows the node exclusion distribution across 224 multi-node training sessions. The distribution is concentrated: the top 3 most-excluded nodes (gpu074, gpu119, gpu086) account for over 50\% of all exclusions, while most nodes have exclusion rates below 5\%.

Over the analysis period, the measured temporal occupancy of 60-node training sessions was approximately 96.6\%. The longest session ran for 222.9 hours (9.3 days), and the top 5 sessions each ran continuously for more than 3.6 days.

\begin{figure}[H]
  \centering
  \includegraphics[width=\linewidth]{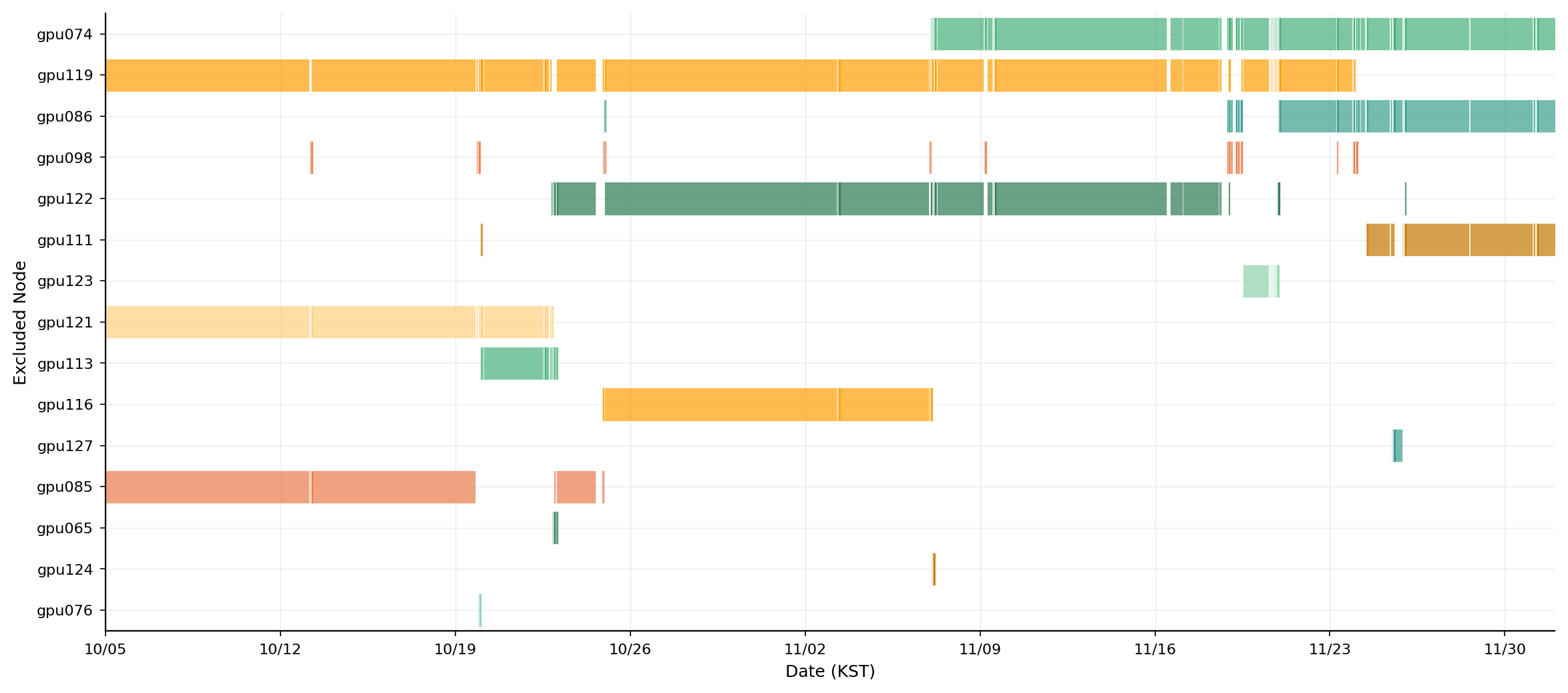}
  \caption{60-node training session exclusion timeline for the top 15 nodes. Each bar represents the duration during which the node was excluded from a training session. Compared with Figure~\ref{fig:single-node-reservation}, most exclusions overlap temporally with deliberate single-node occupancy.}
  \label{fig:node-exclusion-timeline}
\end{figure}

\begin{figure}[H]
  \centering
  \includegraphics[width=\linewidth]{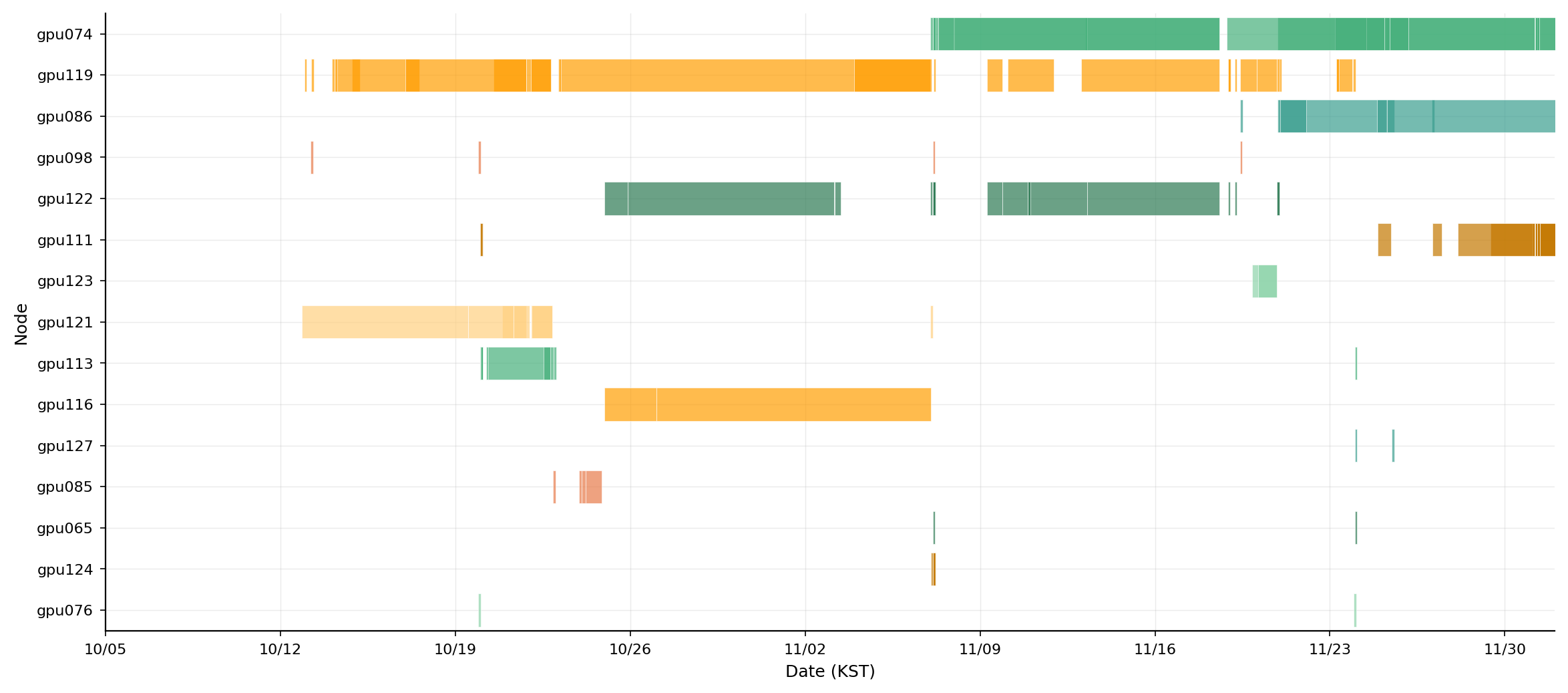}
  \caption{Single-node session occupancy timeline for the same 15 nodes. The pattern shows operators allocating single-node sessions to problem nodes to deliberately exclude them from 60-node training scheduling. For gpu074 (100\%), gpu086 (97\%), and gpu116 (99.6\%), nearly all of the 60-node-training exclusion time overlaps with single-node occupancy.}
  \label{fig:single-node-reservation}
\end{figure}

To analyze the causes of node exclusion, we computed the fraction of 60-node-training exclusion time that overlaps with single-node session allocation on the same node (Figure~\ref{fig:node-exclusion-timeline}, \ref{fig:single-node-reservation}). This ratio serves as an indicator distinguishing whether a node was deliberately isolated by an operator (single-node session occupancy) or naturally not selected as a consequence of the scheduler choosing 60 out of 63 nodes.

Many of the top-excluded nodes correspond to deliberate isolation. For gpu074 (100\%), gpu086 (97\%), gpu116 (99.6\%), and gpu113 (92\%), nearly all the exclusion time overlaps with single-node occupancy, the result of operators explicitly isolating these nodes by assigning single-node sessions out of concern for performance degradation (communication delays, reduced training speed, etc.). gpu119 (69\%, with absolute overlap of 793 hours) and gpu122 (72\%, 447 hours) show somewhat lower ratios, but their absolute overlap times are substantial, classifying them as nodes frequently subject to deliberate isolation.

In contrast, gpu085 (4\% of 393 excluded hours) and gpu098 (2\% of 20 excluded hours) barely overlap with single-node occupancy, suggesting they were not deliberately isolated but naturally not selected as a consequence of the scheduler choosing 60 out of 63 nodes.

In summary, the top 3 nodes (gpu074, gpu119, gpu086) account for over 50\% of all exclusions in a concentrated distribution, and many of these correspond to operators' deliberate exclusion (explicit isolation via single-node session occupancy). Some nodes (such as gpu085) do not overlap with single-node occupancy and appear to have been naturally excluded as a consequence of the scheduler's random non-selection.

\subsubsection{Auto-Retry Chain Analysis}
\label{sec:auto-retry-chain}

\begin{figure}[H]
  \centering
  \includegraphics[width=0.88\linewidth]{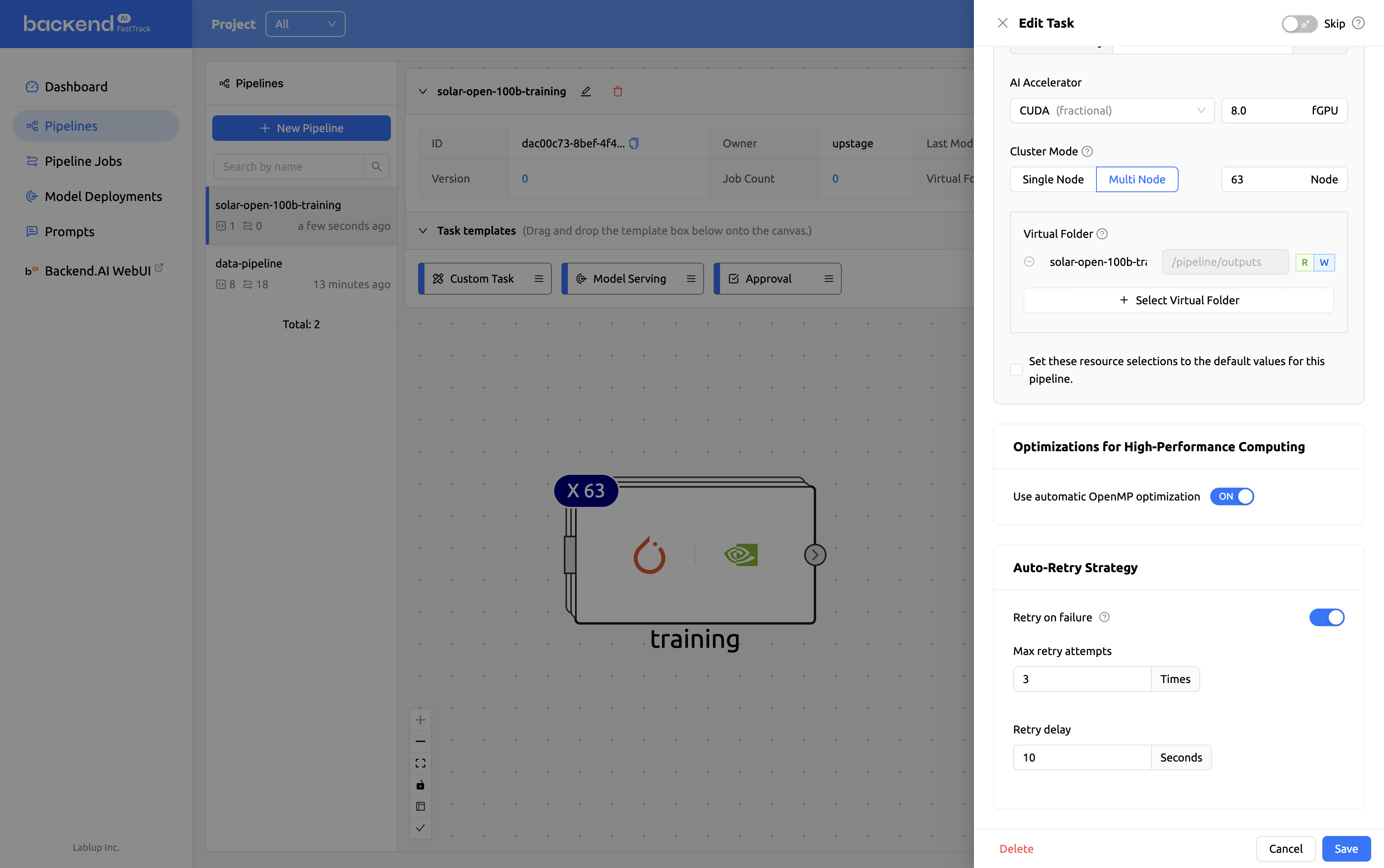}
  \caption{Backend.AI FastTrack auto-retry configuration screen. Operators can configure per-task settings for retry on failure, maximum retry count, and retry delay. Additional options include resource configurations such as GPU count, node count, and storage mounts.}
  \label{fig:fasttrack-auto-retry}
\end{figure}

The auto-retry analysis evaluates how quickly and how reliably recovery proceeds after failure. Backend.AI FastTrack exposes auto-retry controls at the task level (Figure~\ref{fig:fasttrack-auto-retry}) through three parameters: whether retry is enabled, the maximum retry count, and the retry delay. During the observation period, retry was enabled and the delay was set to approximately 10 minutes. The following results quantify the operational effect of that configuration.

From 73 days of operational logs, sessions consecutively executed under the same task name were grouped as auto-retry ``chains.'' Twelve such chains (73 attempts in total, 61 retries) were identified, with results shown in Table~\ref{tab:auto-retry-chains}.

\begin{table}[H]
\centering
\small
\caption{Result classification of 12 auto-retry chains.}
\label{tab:auto-retry-chains}
\begin{tabular}{@{}llr@{}}
\toprule
\textbf{Result} & \textbf{Description} & \textbf{Count} \\
\midrule
SUCCESS & Training reached after retry & 4 \\
FAIL (after training) & First attempt reached training, all subsequent retries failed & 3 \\
FAIL (start failure) & Failed from the start, all retries failed & 5 \\
\bottomrule
\end{tabular}
\end{table}

\begin{figure}[H]
  \centering
  \includegraphics[width=\linewidth]{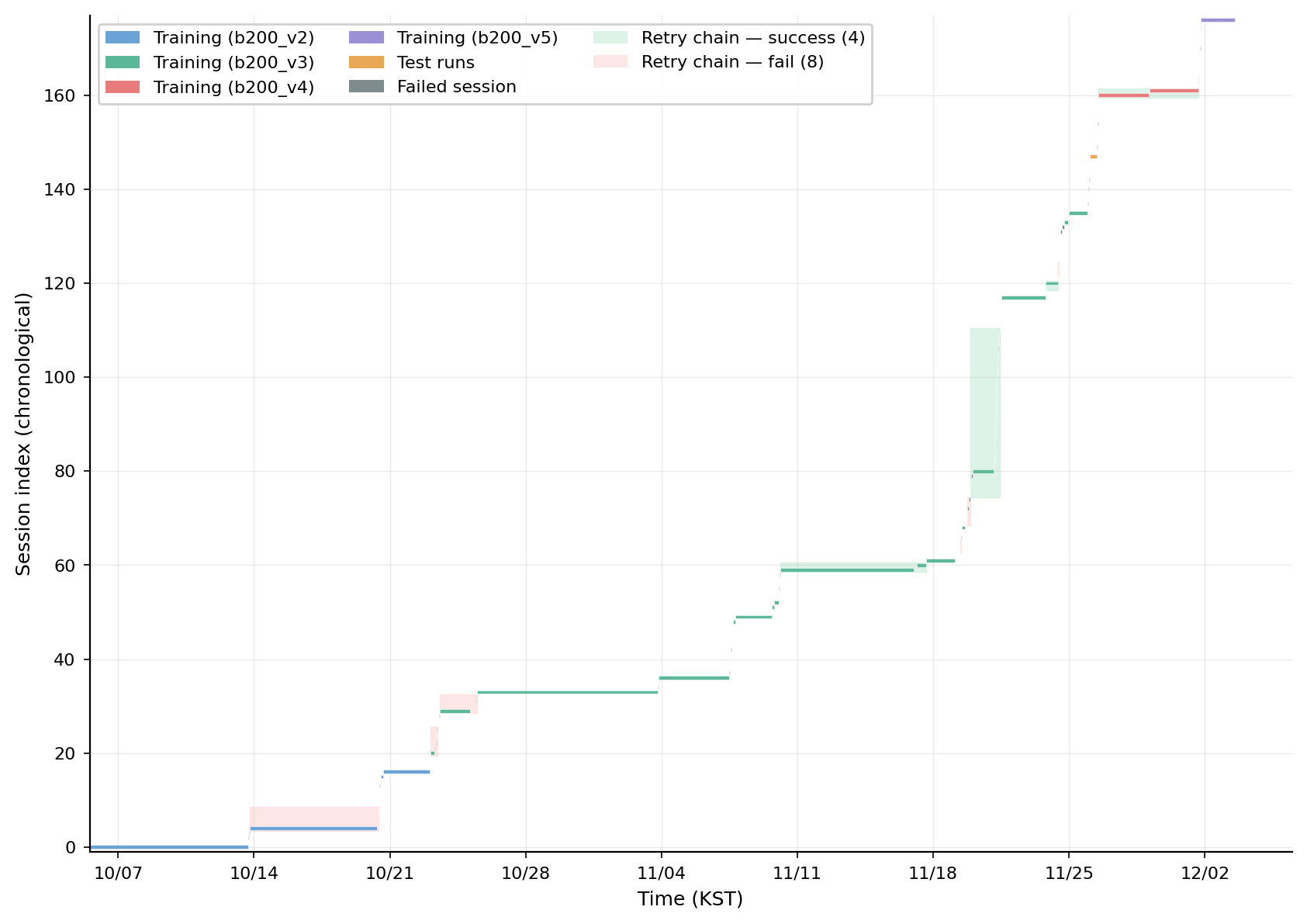}
  \caption{73-day session timeline. The x-axis is time; the y-axis is the order of session start times. Each bar represents a single 60-node session, and colors distinguish training versions (b200\_v2--v5). Background highlights indicate auto-retry chains: green for chains that successfully resumed training after retries (4), pink for chains that failed (8).}
  \label{fig:auto-retry-session-overview}
\end{figure}

Figure~\ref{fig:auto-retry-session-overview} shows the chronological timeline of all sessions.

\subsubsection{Retry Interval Predictability}
\label{sec:retry-interval}

\begin{figure}[H]
  \centering
  \includegraphics[width=\linewidth]{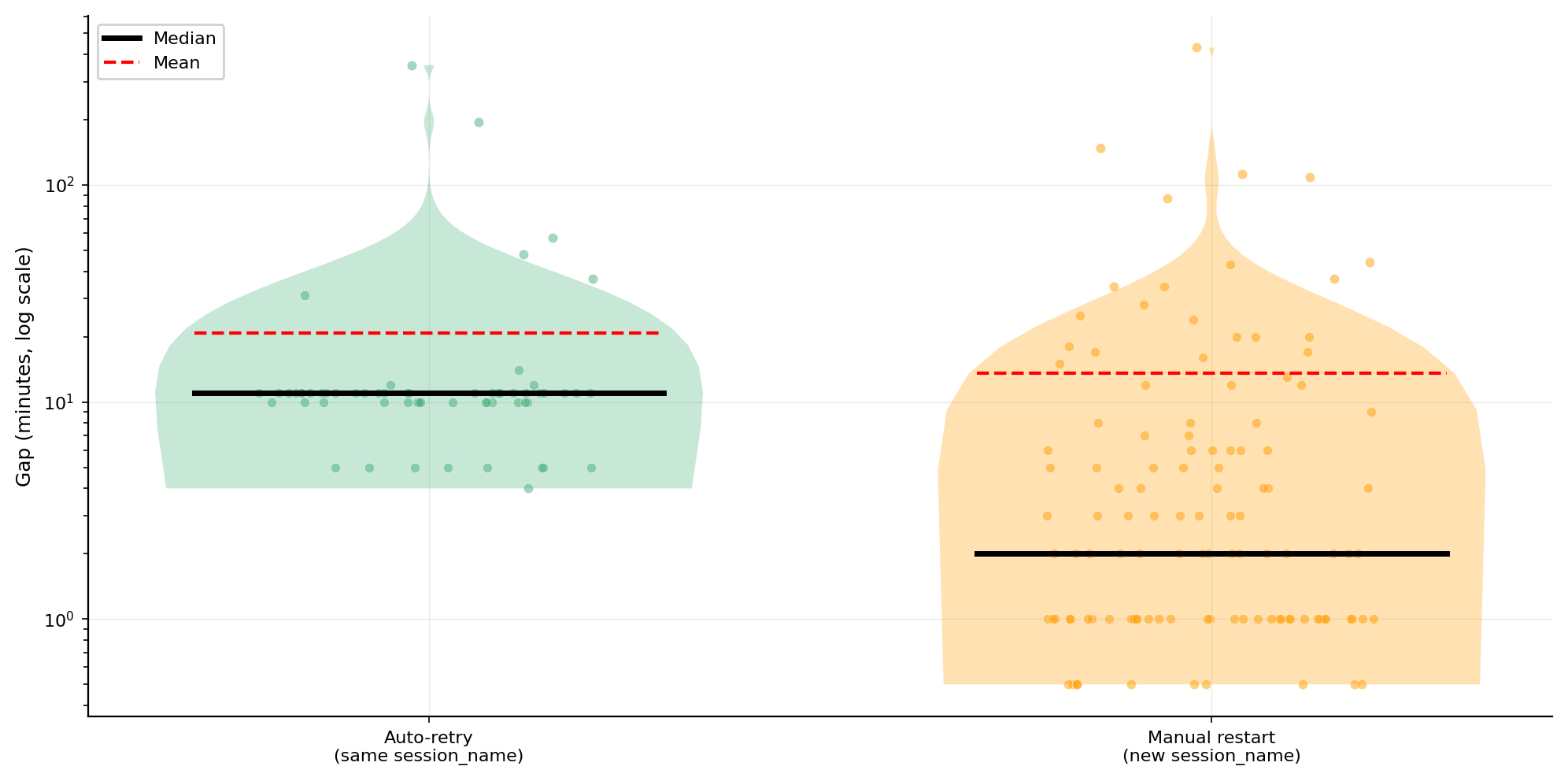}
  \caption{Comparison of inter-session gap distributions between auto-retry and manual restart. Auto-retry has median 11 minutes, IQR 10--11 minutes, corresponding to the FastTrack retry delay setting. Manual restart has median 2 minutes but a range of 0--430 minutes, unpredictable depending on response timing.}
  \label{fig:auto-retry-gap-comparison}
\end{figure}

FastTrack's configured retry delay produces a highly regular restart cadence (Figure~\ref{fig:auto-retry-gap-comparison}). Auto-retry inter-session gaps have a median of 11 minutes with IQR 10--11 minutes, which matches the 10-minute retry delay plus teardown and restart overhead. Manual restarts have a shorter median of 2 minutes but a far wider range of 0--430 minutes, making them operationally unpredictable. The contrast is especially important at night and on weekends, when human response may be delayed but auto-retry continues to act on a fixed schedule.

\subsubsection{Success Rate Comparison and Downtime Reduction}
\label{sec:success-rate-downtime}

Auto-retry improves recovery success for transient failures, although its benefits are bounded by the structure of the underlying failure. The chain success rate---the fraction of retry sequences under the same task name that reached training at least once---was 33.3\% (4 of 12 chains). By comparison, only 12.5\% of manually started individual sessions (13 of 104) reached training, making the chain success rate approximately 2.7$\times$ higher. Chains are naturally advantaged because they include multiple attempts, but the gap still reflects the structural value of automated retries. Three of the 4 successful chains recovered after a single retry. One of them involved XID~94 (ECC error), which Table~\ref{tab:xid-classification} classifies as \texttt{RESTART\_APP}; this case shows that auto-retry can restore progress without operator intervention.

\begin{figure}[H]
  \centering
  \includegraphics[width=\linewidth]{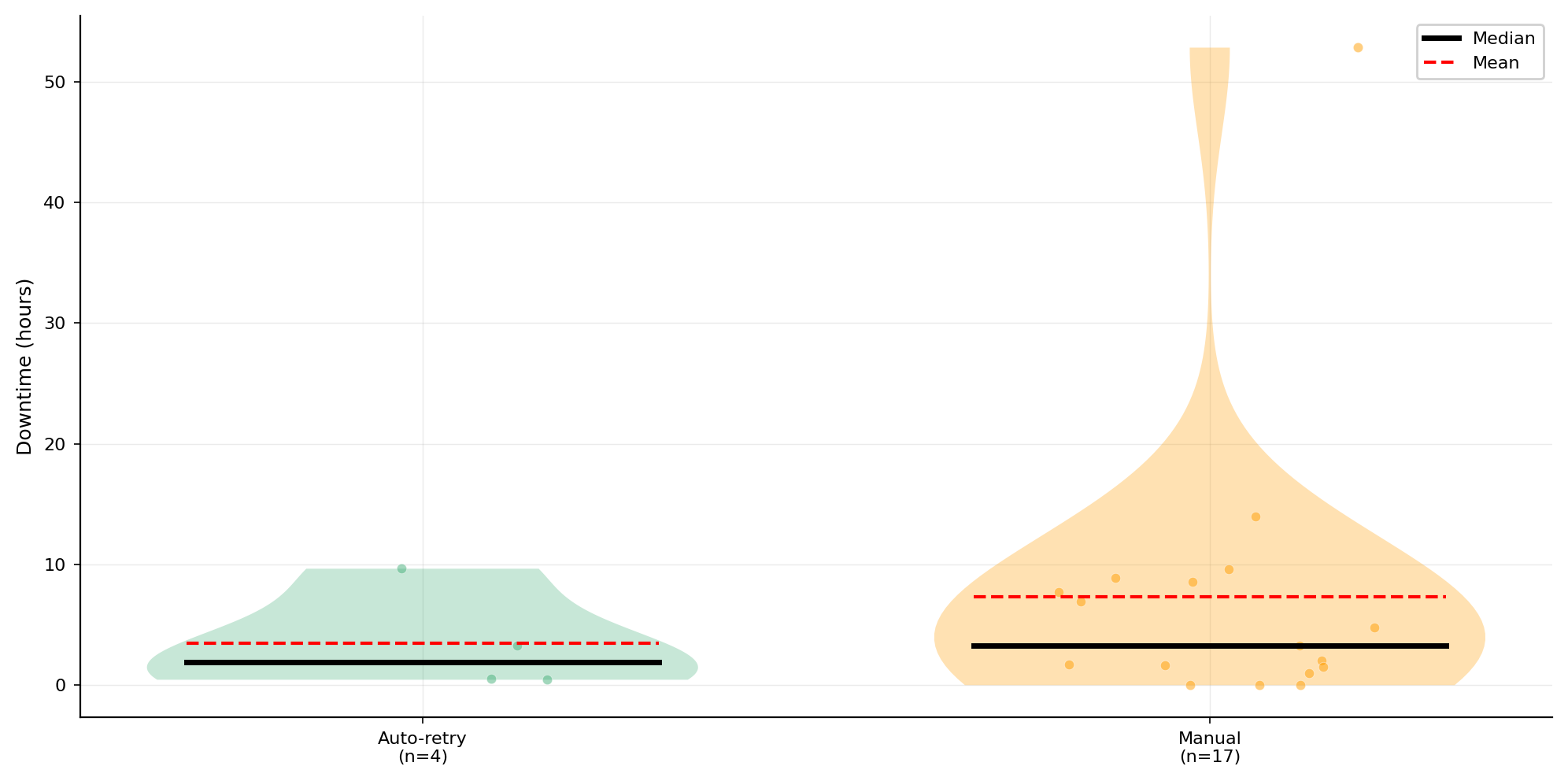}
  \caption{Downtime comparison between auto-retry and manual recovery (21 recovery episodes between training sessions). Auto-retry recoveries (green): median 1.9 hours. Manual recoveries (orange): median 3.3 hours. Black solid line = median, red dashed line = mean.}
  \label{fig:auto-retry-downtime}
\end{figure}

This improvement in success rate translates directly to downtime reduction. Analyzing 21 recovery episodes across 22 training sessions (Figure~\ref{fig:auto-retry-downtime}), the 4 episodes where auto-retry restored training had a median downtime of 1.9 hours, compared to 3.3 hours for 17 manual recovery episodes---a difference of approximately 1.8$\times$. The large variance in manual recovery (0--53 hours) reflects cases where failures occurred during nighttime or weekends when immediate response was difficult.

\subsubsection{Limitations and Future Improvements}
\label{sec:auto-retry-limits}

Eight of 12 chains (67\%) ultimately failed, and most of these failures were caused by software- or network-level issues (for example, NCCL communication errors) that simple restarts could not resolve.

Additionally, auto-retry episodes with long downtimes (9.65 hours, 3.25 hours) were caused by infrastructure-level problems rather than limitations of the auto-retry mechanism itself. After hardware replacement, GPU licenses were not renewed, preventing nodes from joining the available resource pool, which caused retries to fail for hours as the 60-node requirement could not be met. This issue was subsequently resolved by switching to a floating license model.

The retry cost of failed chains was approximately 35 GPU-hours (2.7\% of total training time). In particular, one chain failed 30 consecutive times after 25.4 hours of successful training, illustrating that repeated retries under the same conditions without resolving the underlying problem only consume GPU-hours.

This analysis suggests the following improvement directions:

\begin{itemize}
  \item Exponential backoff: increasing retry intervals progressively (10 min $\rightarrow$ 20 min $\rightarrow$ 40 min) to reduce resource consumption in later retries while maintaining fast initial recovery for transient failures.
  \item XID-based branching: differentiating retry strategies by resolution type from Table~\ref{tab:xid-classification}. \texttt{RESTART\_APP} types (XID~31, 43, 94) retry immediately; \texttt{RESET\_GPU} types (XID~119, 145, 149) retry after GPU reset; \texttt{CONTACT\_SUPPORT} types (XID~79) halt retries and notify operators.
  \item Priority-based session preemption: with only 3 spare nodes (Section~\ref{sec:node-exclusion}), if single-node sessions occupy them, 60-node gang scheduling cannot meet its requirements and auto-retry can be delayed. Granting higher priority to multi-node training and automatically preempting lower-priority single-node sessions during retries, or expanding the spare node pool, could improve availability.
\end{itemize}

\section{Limitations}
\label{sec:evaluation}

This report analyzes operating data from multiple systems, with an emphasis on infrastructure-level failures and recovery behavior. The checkpoint I/O analysis in Section~\ref{sec:checkpoint-io} quantifies checkpoint overhead ($\delta = 18\text{--}31.7$ seconds), lost time per failure (mean 0.98 hours), restart loading time (mean 33 minutes), and NFS/RPC request queueing. Evaluating the effect of these infrastructure events on training throughput and convergence time would require additional instrumentation inside the training framework.

The precursor analysis in Section~\ref{sec:precursor-analysis} is also retrospective. Real-time deployment would require separate validation of the false-positive distribution and the operational burden imposed by alerts. The 10 analyzed failure cases further limit statistical power.

Finally, the auto-retry analysis covers only 12 chains (73 attempts). The results are sufficient to reveal structural properties of recovery behavior, but they remain limited in sample size.

\section{Related Work}
\label{sec:related-work}

\paragraph{GPU cluster scheduling.}
GPU cluster scheduling has progressed from attained-service-based priority (Tiresias~\cite{gu2019tiresias}) and introspective time-slicing (Gandiva~\cite{xiao2018gandiva}) to goodput-adaptive systems such as Pollux~\cite{qiao2021pollux} and Sia~\cite{subramanya2023sia}, and more recently to work on fairness~\cite{zheng2023shockwave}, network topology~\cite{rajasekaran2024cassini}, geo-distributed scheduling~\cite{choudhury2024mast}, and cloud resource management~\cite{wang2024dlrover}. These systems typically optimize JCT or goodput through workload-adaptive control. By contrast, Sokovan emphasizes predictable scheduling latency through hint-based polling and does not currently perform per-workload adaptive optimization.

\paragraph{Distributed training systems.}
Distributed training research has focused primarily on parallelization strategy, including model parallelism (Megatron-LM~\cite{shoeybi2019megatron, narayanan2021megatron}), pipeline parallelism (GPipe~\cite{huang2019gpipe}), memory-efficient sharding (DeepSpeed ZeRO~\cite{rajbhandari2020zero}, PyTorch FSDP~\cite{zhao2023fsdp}), automatic strategy selection (Alpa~\cite{zheng2022alpa}), and 10{,}000+ GPU scaling (MegaScale~\cite{jiang2024megascale}). Backend.AI operates at a different layer: it manages session lifecycle and resource allocation rather than parallelization itself.

\paragraph{Failure recovery and checkpointing.}
Checkpoint-based recovery remains fundamental to long-running training. Prior work spans frequency optimization (CheckFreq~\cite{mohan2021checkfreq}), predictive checkpointing~\cite{gupta2024jit}, in-memory checkpoints (Gemini~\cite{wang2023gemini}, ByteCheckpoint~\cite{wan2025bytecheckpoint}), checkpoint-free recovery through pipeline templates (Oobleck~\cite{jang2023oobleck}), and redundant computation (Bamboo~\cite{thorpe2023bamboo}). Training recovery systems include elastic spot-VM training (Varuna~\cite{athlur2022varuna}), fast failure detection (TRANSOM~\cite{wu2023transom}), and MoE-specific recovery (Lazarus~\cite{wu2024lazarus}), the last of which targets the same architectural class as the workload studied here. These systems primarily optimize checkpointing behavior or recovery inside the training stack itself. Our auto-retry mechanism instead performs session restart and resource reallocation at the orchestration layer while delegating checkpoint creation and restoration to the training framework. Section~\ref{sec:restart-loading} quantifies the resulting restart loading time.

\section{Conclusion}

This report analyzed failure precursor detection, checkpoint I/O behavior, and automated recovery on a 63-node B200 production GPU cluster using 55 days of monitoring data and 73 days of operational logs.

\subsection{Summary of Key Findings}

The three analyses in Section~\ref{sec:case-studies} yield four findings, which we distill below into three broader principles. All three depend on the cross-organizational operational setting described in Section~\ref{sec:cross-org-setting}: without a shared metric pipeline across organizational boundaries, the production-scale phenomena summarized here would not have been directly observable.

\paragraph{Failures are a structural characteristic of large-scale training.} The mathematical relationship between cluster scale and failure frequency, along with operational evidence (419 interruptions over 54 days on 16K GPUs; concentrated node exclusion patterns on a 63-node cluster), confirms that hardware failures every few hours are a fundamental characteristic of large-scale training.

\paragraph{Training workloads require session-level abstraction.} Container orchestration assumes stateless, short-lived processes. Training workloads are stateful and long-running, requiring an abstraction that tracks checkpoint progress and enables resumption rather than restart. Backend.AI's session abstraction decouples training progress from container lifecycle (Section~\ref{sec:session-recovery}).

\paragraph{GPU scheduling and storage must be co-designed.} CPU-centric resource models fail to capture GPU topology or all-or-nothing allocation requirements. The storage I/O bottleneck observed in our cross-organizational setting (Section~\ref{sec:cross-org-setting}) illustrates this: provisioning GPU capacity without matching storage processing paths creates performance cliffs that manifest only at operational scale. Full-stack profiling of checkpoint I/O (Section~\ref{sec:rpc-bottleneck}) shows average throughput of 21.5\% of the storage's maximum read bandwidth during restart loading and 16.0\% of the write bandwidth during checkpoint-save bursts, with concurrent increases in NFS/RPC request queueing and transport-layer backlog. Improvement should therefore focus not only on network-link bandwidth, but also on maintaining sufficient parallelism as checkpoint I/O traverses the shared NFS path. Sokovan provides the GPU-side solution through GPU-first allocation combined with gang scheduling.

\subsection{Future Work}
\label{sec:future-work}

Several limitations of this study stem from missing instrumentation during the observation period. Future training campaigns should therefore add the following measurements.

\paragraph{Training efficiency metrics.}
The current analysis measures infrastructure-level figures (checkpoint intervals, restart times, failure rates) but lacks metrics internal to the training framework. Logging per-iteration throughput (tokens/sec) would enable MFU calculation and direct quantification of how infrastructure events---failures, restarts, node replacements---impact effective training progress. This can be collected through a simple configuration change to the training framework logger before training begins.

\paragraph{NFS/RPC path optimization.}
Section~\ref{sec:rpc-bottleneck} shows that the strongest checkpoint I/O signals appear in NFS/RPC request queueing and transport-layer backlog. During saves, WRITE requests spend 1.89 seconds of their 2.03-second average request time in queue (93.1\%); during loads, READ requests spend 54.2\,ms of their 78.2\,ms average request time in queue (57.0\%). Future runs should collect NFS mount options (\texttt{nconnect}, rsize/wsize), readahead, client- and server-side queue metrics, NFS frontend metrics, and backend storage write metrics together, so that the layer where queueing forms can be separated. A dedicated high-throughput checkpoint I/O path should also be evaluated.

\paragraph{Precursor-based predictive failure management.}
In the statistical multi-signal detection of Section~\ref{sec:precursor-analysis}, the low pre-XID detection rate is the principal limitation. In many failures on this cluster, signals do not deteriorate gradually but emerge abruptly at the XID time point, which makes pre-XID detection difficult. To address this, we are pursuing follow-on ML modeling that learns multivariate time-series patterns and changes in cross-metric correlations. For operational deployment, we must also verify that the false-positive level holds under real-time inference and design an integrated path through which detection results inform auto-retry decisions.

\paragraph{Intelligent resource adjustment and log analysis.}
Future work includes exploring an automated operations design that combines automatic resource adjustment upon OOM~\cite{kang2026oomrecovery}, detection of resource over-allocation, and integrated analysis across heterogeneous logs.

\paragraph{FP8 and reduced-precision training.}
Reduced-precision formats such as FP8 and MXFP8 promise throughput improvements but introduce failure modes in both the software stack and numerical stability. NVIDIA cuDNN releases document numerous FP8-related defects~\cite{nvidia2024cudnn}, Fishman et al.~\cite{fishman2025scalingfp8} showed catastrophic instability after approximately 200 billion tokens, and Lee et al.~\cite{lee2024fp8back} demonstrated the lack of general robustness in current FP8 methods. The Solar Open project adopted FP8~+~bfloat16 mixed precision on the same B200 cluster~\cite{upstage2026solar}. From an infrastructure perspective, developing an automated failure attribution mechanism that distinguishes whether training divergence originates from cuDNN bugs, numerical limitations, or hardware defects remains an open challenge.

\section*{Acknowledgments}

This work was conducted as part of Korea's Sovereign AI Project (GPU Track), led by the Ministry of Science and ICT (MSIT) and supported by the National IT Industry Promotion Agency (NIPA) (PJT-25-080041).

The storage I/O debugging case study in Section~\ref{sec:storage-debugging} was made possible by collaboration across multiple organizations. We thank SKT for cloud infrastructure and operational support, NVIDIA Korea for hardware expertise and driver-level diagnostics, VAST Data for storage-system analysis and configuration optimization, and Upstage for sharing workload characteristics and participating in joint debugging sessions.

We also thank the open-source communities that provide the tools and frameworks on which Backend.AI is built, including PyTorch, NCCL, and the Linux kernel networking stack.

\section*{Data Availability}

The operational data analyzed in this report---including Prometheus time-series metrics, node exclusion logs, auto-retry records, and GPU utilization traces---was collected from a production cluster operated under Korea's Sovereign AI Project and contains proprietary workload information. These datasets cannot be released because of contractual and confidentiality constraints.
The Backend.AI platform itself is available as open-source software at \url{https://github.com/lablup/backend.ai}. The \texttt{all-smi} monitoring tool is available at \url{https://github.com/lablup/all-smi}.
Aggregate statistics sufficient to reproduce the analyses presented in this report are provided in the tables and figures within the main text.

\bibliographystyle{unsrt}
\bibliography{references}

@inproceedings{xiao2018gandiva,
  author    = {Xiao, Wencong and Bhardwaj, Romil and Ramjee, Ramachandran and Sivathanu, Muthian and Kwatra, Nipun and Han, Zhenhua and Patel, Pratyush and Peng, Xuan and Zhao, Hanyu and Zhang, Quanlu and Yang, Fan and Zhou, Lidong},
  title     = {Gandiva: Introspective Cluster Scheduling for Deep Learning},
  booktitle = {Proceedings of the 13th USENIX Symposium on Operating Systems Design and Implementation (OSDI)},
  year      = {2018},
  pages     = {595--610},
  publisher = {USENIX Association}
}

@inproceedings{gu2019tiresias,
  author    = {Gu, Juncheng and Chowdhury, Mosharaf and Shin, Kang G. and Zhu, Yibo and Jeon, Myeongjae and Qian, Junjie and Liu, Hongqiang and Guo, Chuanxiong},
  title     = {Tiresias: A GPU Cluster Manager for Distributed Deep Learning},
  booktitle = {Proceedings of the 16th USENIX Symposium on Networked Systems Design and Implementation (NSDI)},
  year      = {2019},
  pages     = {485--500},
  publisher = {USENIX Association}
}

@inproceedings{qiao2021pollux,
  author    = {Qiao, Aurick and Choe, Sang Keun and Subramanya, Suhas Jayaram and Neiswanger, Willie and Ho, Qirong and Zhang, Hao and Ganger, Gregory R. and Xing, Eric P.},
  title     = {Pollux: Co-adaptive Cluster Scheduling for Goodput-optimized Deep Learning},
  booktitle = {Proceedings of the 15th USENIX Symposium on Operating Systems Design and Implementation (OSDI)},
  year      = {2021},
  pages     = {1--18},
  publisher = {USENIX Association}
}

@inproceedings{jeon2019analysis,
  author = {Myeongjae Jeon and Shivaram Venkataraman and Amar Phanishayee and Junjie Qian and Wencong Xiao and Fan Yang},
  title = {Analysis of {Large-Scale} {Multi-Tenant} {GPU} Clusters for {DNN} Training Workloads},
  booktitle = {2019 USENIX Annual Technical Conference (USENIX ATC 19)},
  year = {2019},
  isbn = {978-1-939133-03-8},
  address = {Renton, WA},
  pages = {947--960},
  url = {https://www.usenix.org/conference/atc19/presentation/jeon},
  publisher = {USENIX Association},
  month = jul
}

@inproceedings{subramanya2023sia,
  author    = {Subramanya, Suhas Jayaram and Arfeen, Daiyaan and Lin, Shouxu and Qiao, Aurick and Jia, Zhihao and Ganger, Gregory R.},
  title     = {Sia: Heterogeneity-aware, goodput-optimized ML-cluster scheduling},
  booktitle = {Proceedings of the 29th ACM Symposium on Operating Systems Principles (SOSP)},
  year      = {2023},
  pages     = {642--657},
  publisher = {ACM},
  doi       = {10.1145/3600006.3613175}
}

@inproceedings{rajasekaran2024cassini,
  author    = {Rajasekaran, Sudarsanan and Ghobadi, Manya and Akella, Aditya},
  title     = {CASSINI: Network-Aware Job Scheduling in Machine Learning Clusters},
  booktitle = {Proceedings of the 21st USENIX Symposium on Networked Systems Design and Implementation (NSDI)},
  year      = {2024},
  publisher = {USENIX Association}
}

@misc{nvidia2026dcgmexporter,
  author       = {{NVIDIA Corporation}},
  title        = {DCGM-Exporter: {NVIDIA} {GPU} Monitoring Tool for {Prometheus}},
  year         = {2026},
  howpublished = {\url{https://github.com/NVIDIA/dcgm-exporter}},
  note         = {Accessed: 2026-02-15}
}

@misc{nvidia2025gpumemerrormgmt,
  author       = {{NVIDIA Corporation}},
  title        = {{NVIDIA} {GPU} Memory Error Management},
  year         = {2025},
  howpublished = {\url{https://docs.nvidia.com/deploy/a100-gpu-mem-error-mgmt/index.html}},
  note         = {v590. Accessed: 2026-03-24}
}

@misc{prometheus2026nodeexporter,
  author       = {{Prometheus Authors}},
  title        = {Node Exporter: {Prometheus} Exporter for Hardware and {OS} Metrics},
  year         = {2026},
  howpublished = {\url{https://github.com/prometheus/node_exporter}},
  note         = {Accessed: 2026-02-15}
}

@misc{nvidia2026xiderrors,
  author       = {{NVIDIA Corporation}},
  title        = {Analyzing {XID} Errors --- {GPU} Deployment and Management Documentation},
  year         = {2026},
  howpublished = {\url{https://docs.nvidia.com/deploy/xid-errors/analyzing-xid-catalog.html}},
  note         = {Accessed: 2026-02-15}
}

@misc{lablup2025allsmi,
  author       = {{Lablup Inc.}},
  title        = {all-smi: Cross-Platform AI Accelerator Monitoring Tool},
  year         = {2025},
  howpublished = {\url{https://github.com/lablup/all-smi}},
  note         = {Accessed: 2026-02-05}
}

@misc{shin2023sokovan,
  author       = {Shin, Jeongkyu and Kim, Joongi},
  title        = {Sokovan: Container Orchestrator for Accelerated {AI/ML} Workloads and Massive-scale {GPU} Computing},
  year         = {2023},
  howpublished = {Presented at OpenInfra Summit Vancouver},
  month        = jun,
  note         = {Slides available at \url{https://www.backend.ai/ko/video/2023-06-11-openinfra-summit}}
}

@inproceedings{fishman2025scalingfp8,
  author    = {Fishman, Maxim and Chmiel, Brian and Banner, Ron and Soudry, Daniel},
  title     = {Scaling {FP8} Training to Trillion-Token {LLMs}},
  booktitle = {Proceedings of the Thirteenth International Conference on Learning Representations (ICLR)},
  year      = {2025},
  note      = {Spotlight. arXiv:2409.12517}
}

@article{lee2024fp8back,
  author  = {Lee, Joonhyung and Bae, Jeongin and Kim, Byeongwook and Kwon, Se Jung and Lee, Dongsoo},
  title   = {To {FP8} and Back Again: Quantifying the Effects of Reducing Precision on {LLM} Training Stability},
  journal = {arXiv preprint arXiv:2405.18710},
  year    = {2024}
}

@misc{nvidia2024cudnn,
  author       = {{NVIDIA Corporation}},
  title        = {cu{DNN} Backend Release Notes},
  year         = {2024},
  howpublished = {\url{https://docs.nvidia.com/deeplearning/cudnn/backend/latest/release-notes.html}},
  note         = {Cumulative release notes covering cuDNN 9.x; multiple FP8/MXFP8/NVFP4-related fixes across versions. Accessed: 2026-04-21}
}

@article{young1974first,
  author  = {Young, John W.},
  title   = {A First Order Approximation to the Optimum Checkpoint Interval},
  journal = {Communications of the ACM},
  volume  = {17},
  number  = {9},
  pages   = {530--531},
  year    = {1974},
  doi     = {10.1145/361147.361115}
}

@article{daly2006higher,
  author  = {Daly, John T.},
  title   = {A Higher Order Estimate of the Optimum Checkpoint Interval for Restart Dumps},
  journal = {Future Generation Computer Systems},
  volume  = {22},
  number  = {3},
  pages   = {303--312},
  year    = {2006},
  doi     = {10.1016/j.future.2004.11.016}
}

@inproceedings{mohan2021checkfreq,
  author    = {Mohan, Jayashree and Phanishayee, Amar and Chidambaram, Vijay},
  title     = {CheckFreq: Frequent, Fine-Grained DNN Checkpointing},
  booktitle = {Proceedings of the 19th USENIX Conference on File and Storage Technologies (FAST)},
  year      = {2021},
  pages     = {203--216},
  publisher = {USENIX Association}
}

@inproceedings{gupta2024jit,
  author    = {Gupta, Tanmaey and Krishnan, Sanjeev and Kumar, Rituraj and Vijeev, Abhishek and Gulavani, Bhargav and Kwatra, Nipun and Ramjee, Ramachandran and Sivathanu, Muthian},
  title     = {Just-In-Time Checkpointing: Low Cost Error Recovery from Deep Learning Training Failures},
  booktitle = {Proceedings of the Nineteenth European Conference on Computer Systems (EuroSys)},
  year      = {2024},
  pages     = {1110--1125},
  publisher = {ACM},
  doi       = {10.1145/3627703.3650085}
}

@article{wang2024dlrover,
  author    = {Wang, Qinlong and Lan, Tingfeng and Tang, Yinghao and Huang, Ziling and Du, Yiheng and Zhang, Haitao and Sha, Jian and Lu, Hui and Zhou, Yuanchun and Zhang, Ke and Tang, Mingjie},
  title     = {{DLRover-RM}: Resource Optimization for Deep Recommendation Models Training in the Cloud},
  journal   = {Proceedings of the VLDB Endowment},
  volume    = {17},
  year      = {2024},
  publisher = {VLDB Endowment},
  doi       = {10.14778/3685800.3685832}
}

@inproceedings{kokolis2025revisiting,
  author    = {Kokolis, Apostolos and Kuchnik, Michael and Hoffman, John and Kumar, Adithya and Malani, Parth and Ma, Faye and DeVito, Zachary and Sengupta, Shubho and Saladi, Kalyan and Wu, Carole-Jean},
  title     = {Revisiting Reliability in Large-Scale Machine Learning Research Clusters},
  booktitle = {Proceedings of the 2025 IEEE International Symposium on High-Performance Computer Architecture (HPCA)},
  year      = {2025},
  publisher = {IEEE}
}

@article{grattafiori2024llama3,
  author    = {Grattafiori, Aaron and Dubey, Abhimanyu and others},
  title     = {The Llama 3 Herd of Models},
  journal   = {arXiv preprint arXiv:2407.21783},
  year      = {2024}
}

@techreport{upstage2026solar,
  author      = {{Upstage Solar Team}},
  title       = {Solar Open Technical Report},
  institution = {Upstage},
  year        = {2026},
  month       = jan,
  type        = {Technical Report},
  url         = {https://arxiv.org/abs/2601.07022},
  note        = {arXiv:2601.07022. 102B bilingual MoE (12B active) trained on 20T tokens. Also available at \url{https://huggingface.co/upstage/Solar-Open-100B}}
}

@misc{erben2024hardware,
  title={Hardware Failures Won't Limit {AI} Scaling},
  author={Alexander Erben and Ege Erdil},
  year={2024},
  howpublished={Epoch AI},
  url={https://epoch.ai/blog/hardware-failures-wont-limit-ai-scaling}
}

@techreport{nvidia2024blackwellbrief,
  author       = {{NVIDIA Corporation}},
  title        = {{NVIDIA} {Blackwell} Architecture Technical Brief: Powering the New Era of Generative {AI} and Accelerated Computing},
  institution  = {NVIDIA Corporation},
  year         = {2024},
  type         = {Technical Brief},
  note         = {Version 1.1. Per Blackwell GPU: up to 192~GB HBM3e, up to 8~TB/s. Accessed: 2026-04-21}
}

@misc{nvidia2025superpodb200,
  author       = {{NVIDIA Corporation}},
  title        = {{DGX} {SuperPOD}: Next Generation Scalable Infrastructure for {AI} Leadership Reference Architecture Featuring {DGX} {B200}},
  year         = {2025},
  howpublished = {\url{https://docs.nvidia.com/dgx-superpod/reference-architecture-scalable-infrastructure-b200/latest/}},
  note         = {Document RA-11334-001. Accessed: 2026-02-18}
}

@inproceedings{jiang2024megascale,
  author    = {Jiang, Ziheng and Lin, Haibin and Zhong, Yinmin and Huang, Qi and Chen, Yangrui and Zhang, Zhi and Peng, Yanghua and Li, Xiang and Xie, Cong and Nong, Shibiao and others},
  title     = {MegaScale: Scaling Large Language Model Training to More Than 10,000 GPUs},
  booktitle = {Proceedings of the 21st USENIX Symposium on Networked Systems Design and Implementation (NSDI)},
  year      = {2024},
  publisher = {USENIX Association}
}

@inproceedings{wang2023gemini,
  author    = {Wang, Zhuang and Jia, Zhen and Zheng, Shuai and Zhang, Zhen and Fu, Xinwei and Ng, T. S. Eugene and Wang, Yida},
  title     = {Gemini: Fast Failure Recovery in Distributed Training with In-Memory Checkpoints},
  booktitle = {Proceedings of the 29th ACM Symposium on Operating Systems Principles (SOSP)},
  year      = {2023},
  pages     = {364--381},
  publisher = {ACM},
  doi       = {10.1145/3600006.3613145}
}

@inproceedings{jang2023oobleck,
  author    = {Jang, Insu and Yang, Zhenning and Zhang, Zhen and Jin, Xin and Chowdhury, Mosharaf},
  title     = {Oobleck: Resilient Distributed Training of Large Models Using Pipeline Templates},
  booktitle = {Proceedings of the 29th ACM Symposium on Operating Systems Principles (SOSP)},
  year      = {2023},
  pages     = {382--395},
  publisher = {ACM},
  doi       = {10.1145/3600006.3613152}
}

@inproceedings{wan2025bytecheckpoint,
  author    = {Wan, Borui and Han, Mingji and Sheng, Yiyao and Peng, Yanghua and Lin, Haibin and Zhang, Mofan and Lai, Zhichao and Yu, Menghan and Zhang, Junda and Song, Zuquan and Liu, Xin and Wu, Chuan},
  title     = {ByteCheckpoint: A Unified Checkpointing System for Large Foundation Model Development},
  booktitle = {Proceedings of the 22nd USENIX Symposium on Networked Systems Design and Implementation (NSDI)},
  year      = {2025},
  publisher = {USENIX Association}
}

@inproceedings{thorpe2023bamboo,
  author    = {Thorpe, John and Zhao, Pengzhan and Eyolfson, Jonathan and Qiao, Yifan and Jia, Zhihao and Zhang, Minjia and Netravali, Ravi and Xu, Guoqing Harry},
  title     = {Bamboo: Making Preemptible Instances Resilient for Affordable Training of Large DNNs},
  booktitle = {Proceedings of the 20th USENIX Symposium on Networked Systems Design and Implementation (NSDI)},
  year      = {2023},
  publisher = {USENIX Association}
}

@article{wu2024lazarus,
  author    = {Wu, Yongji and Qu, Wenjie and Liu, Xueshen and Tao, Tianyang and Qiao, Yifan and Wang, Zhuang and Bai, Wei and Tian, Yuan and Zhang, Jiaheng and Mao, Z. Morley and Lentz, Matthew and Zhuo, Danyang and Stoica, Ion},
  title     = {Lazarus: Resilient and Elastic Training of Mixture-of-Experts Models},
  journal   = {arXiv preprint arXiv:2407.04656},
  year      = {2024}
}

@article{wu2023transom,
  author    = {Wu, Baodong and Xia, Lei and Li, Qingping and Li, Kangyu and Chen, Xu and Guo, Yongqiang and Xiang, Tieyao and Chen, Yuheng and Li, Shigang},
  title     = {{TRANSOM}: An Efficient Fault-Tolerant System for Training {LLMs}},
  journal   = {arXiv preprint arXiv:2310.10046},
  year      = {2023}
}

@article{shoeybi2019megatron,
  author    = {Shoeybi, Mohammad and Patwary, Mostofa and Puri, Raul and LeGresley, Patrick and Casper, Jared and Catanzaro, Bryan},
  title     = {Megatron-LM: Training Multi-Billion Parameter Language Models Using Model Parallelism},
  journal   = {arXiv preprint arXiv:1909.08053},
  year      = {2019}
}

@inproceedings{narayanan2021megatron,
  author    = {Narayanan, Deepak and Shoeybi, Mohammad and Casper, Jared and LeGresley, Patrick and Patwary, Mostofa and Korthikanti, Vijay and Vainbrand, Dmitri and Kasber, Prethvi and Zaharia, Matei and Catanzaro, Bryan},
  title     = {Efficient Large-Scale Language Model Training on GPU Clusters Using Megatron-LM},
  booktitle = {Proceedings of the International Conference for High Performance Computing, Networking, Storage and Analysis (SC)},
  year      = {2021},
  publisher = {ACM}
}

@inproceedings{rajbhandari2020zero,
  author    = {Rajbhandari, Samyam and Rasley, Jeff and Ruwase, Olatunji and He, Yuxiong},
  title     = {ZeRO: Memory Optimizations Toward Training Trillion Parameter Models},
  booktitle = {Proceedings of the International Conference for High Performance Computing, Networking, Storage and Analysis (SC)},
  year      = {2020},
  publisher = {IEEE}
}

@article{zhao2023fsdp,
  author    = {Zhao, Yanli and Gu, Andrew and Varma, Rohan and Luo, Liang and Huang, Chien-Chin and Xu, Min and Wright, Les and Shojanazeri, Hamid and Ott, Myle and Shleifer, Sam and others},
  title     = {PyTorch FSDP: Experiences on Scaling Fully Sharded Data Parallel},
  journal   = {Proceedings of the VLDB Endowment},
  volume    = {16},
  number    = {12},
  pages     = {3848--3860},
  year      = {2023},
  publisher = {VLDB Endowment}
}

@inproceedings{zheng2022alpa,
  author    = {Zheng, Lianmin and Li, Zhuohan and Zhang, Hao and Zhuang, Yonghao and Chen, Zhifeng and Huang, Yanping and Wang, Yida and Xu, Yuanzhong and Zhuo, Danyang and Xing, Eric P. and Gonzalez, Joseph E. and Stoica, Ion},
  title     = {Alpa: Automating Inter- and Intra-Operator Parallelism for Distributed Deep Learning},
  booktitle = {Proceedings of the 16th USENIX Symposium on Operating Systems Design and Implementation (OSDI)},
  year      = {2022},
  pages     = {559--578},
  publisher = {USENIX Association}
}

@inproceedings{huang2019gpipe,
  author    = {Huang, Yanping and Cheng, Youlong and Bapna, Ankur and Firat, Orhan and Chen, Mia Xu and Chen, Dehao and Lee, HyoukJoong and Ngiam, Jiquan and Le, Quoc V. and Wu, Yonghui and Chen, Zhifeng},
  title     = {GPipe: Efficient Training of Giant Neural Networks using Pipeline Parallelism},
  booktitle = {Proceedings of the 33rd International Conference on Neural Information Processing Systems (NeurIPS)},
  year      = {2019},
  pages     = {103--112}
}

@inproceedings{athlur2022varuna,
  author    = {Athlur, Sanjith and Saran, Nitika and Sivathanu, Muthian and Ramjee, Ramachandran and Kwatra, Nipun},
  title     = {Varuna: Scalable, Low-cost Training of Massive Deep Learning Models},
  booktitle = {Proceedings of the Seventeenth European Conference on Computer Systems (EuroSys)},
  year      = {2022},
  pages     = {472--487},
  publisher = {ACM},
  doi       = {10.1145/3492321.3519584}
}

@inproceedings{amaral2017topology,
  author    = {Amaral, Marcelo and Polo, Jord\`{a} and Carrera, David and Seelam, Seetharami and Steinder, Malgorzata},
  title     = {Topology-Aware GPU Scheduling for Learning Workloads in Cloud Environments},
  booktitle = {Proceedings of the International Conference for High Performance Computing, Networking, Storage and Analysis (SC)},
  year      = {2017},
  articleno = {17},
  pages     = {1--12},
  publisher = {ACM},
  doi       = {10.1145/3126908.3126933}
}

@inproceedings{zheng2023shockwave,
  author    = {Zheng, Pengfei and Pan, Rui and Khan, Tarannum and Venkataraman, Shivaram and Akella, Aditya},
  title     = {Shockwave: Fair and Efficient Cluster Scheduling for Dynamic Adaptation in Machine Learning},
  booktitle = {Proceedings of the 20th USENIX Symposium on Networked Systems Design and Implementation (NSDI)},
  year      = {2023},
  publisher = {USENIX Association}
}

@inproceedings{choudhury2024mast,
  author    = {Choudhury, Arnab and Wang, Yang and Pelkonen, Tuomas and Srinivasan, Kutta and Jain, Abha and Lin, Shenghao and David, Delia and Soleimanifard, Siavash and Chen, Michael and Yadav, Abhishek and Tijoriwala, Ritesh and Samoylov, Denis and Tang, Chunqiang},
  title     = {MAST: Global Scheduling of ML Training across Geo-Distributed Datacenters at Hyperscale},
  booktitle = {Proceedings of the 18th USENIX Symposium on Operating Systems Design and Implementation (OSDI)},
  year      = {2024},
  publisher = {USENIX Association}
}

@misc{kang2026oomrecovery,
  author       = {Jungsuk Kang and Joongi Kim},
  title        = {Method and Apparatus for Automatic Recovery of Tasks Using Execution Failure-Based Resource Requirement Adjustment},
  year         = {2026},
  note         = {Korean Patent Application 10-2026-0024429},
  howpublished = {Lablup Inc.}
}

@inproceedings{deng2025minder,
  author    = {Deng, Yangtao and Shi, Xiang and Jiang, Zhuo and Zhang, Xingjian and Zhang, Lei and Zhang, Zhang and Li, Bo and Song, Zuquan and Zhu, Hang and Liu, Gaohong and Li, Fuliang and Wang, Shuguang and Lin, Haibin and Ye, Jianxi and Yu, Minlan},
  title     = {Minder: Faulty Machine Detection for Large-scale Distributed Model Training},
  booktitle = {Proceedings of the 22nd USENIX Symposium on Networked Systems Design and Implementation (NSDI)},
  year      = {2025},
  pages     = {505--521},
  publisher = {USENIX Association}
}

@article{wu2024falcon,
  author    = {Wu, Tianyuan and Wang, Wei and Yu, Yinghao and Yang, Siran and Yang, Wenchao and Duan, Qinkai and Yang, Guodong and Wang, Jiamang and Qu, Lin and Zhang, Liping},
  title     = {{FALCON}: Pinpointing and Mitigating Stragglers for Large-Scale Hybrid-Parallel Training},
  journal   = {arXiv preprint arXiv:2410.12588},
  year      = {2024}
}

@article{lin2025whatif,
  author    = {Lin, Jinkun and Jiang, Ziheng and Song, Zuquan and Zhao, Sida and Yu, Menghan and Wang, Zhanghan and Wang, Chenyuan and Shi, Zuocheng and Shi, Xiang and Jia, Wei and Liu, Zherui and Wang, Shuguang and Lin, Haibin and Liu, Xin and Panda, Aurojit and Li, Jinyang},
  title     = {Understanding Stragglers in Large Model Training Using What-if Analysis},
  journal   = {arXiv preprint arXiv:2505.05713},
  year      = {2025}
}

@article{herault2024survey,
  title     = {A Survey on Checkpointing Strategies: Should We Always Checkpoint \`{a} la {Young}/{Daly}?},
  author    = {Bautista-Gomez, Leonardo and Benoit, Anne and Di, Sheng and Herault, Thomas and Robert, Yves and Sun, Hongyang},
  journal   = {Future Generation Computer Systems},
  volume    = {161},
  pages     = {315--328},
  year      = {2024},
  publisher = {Elsevier},
  doi       = {10.1016/j.future.2024.07.022}
}

\appendix
\section{System Architecture Details}
\label{sec:appendix-architecture}

This appendix describes the detailed implementation of the Backend.AI infrastructure summarized in Section~\ref{sec:design-requirements}. The core design of the Sokovan scheduler (two-level scheduling, NUMA-aware placement, gang scheduling) is covered in Section~\ref{sec:sokovan}; here we describe the health check and storage architectures.

\subsection{Multi-Layer Health Checks}

Table~\ref{tab:health-check-layers} shows Backend.AI's health check architecture.

\begin{table}[H]
\centering
\caption{Multi-layer health check architecture}
\label{tab:health-check-layers}
\resizebox{\linewidth}{!}{%
\begin{tabular}{@{}llr@{}}
\toprule
\textbf{Layer} & \textbf{Mechanism} & \textbf{Timeout / Threshold} \\
\midrule
Infrastructure (etcd) & Periodic liveness probe & 5.0\,s \\
Infrastructure (Valkey/Redis) & Per-component ping & 2.0\,s per component, 5.0\,s total \\
Infrastructure (PostgreSQL) & Periodic liveness probe & 2--5\,s \\
Agent RPC & Manager $\rightarrow$ Agent ping & 5.0\,s \\
Agent Liveness & Heartbeat + sweep & 300\,s timeout, 600\,s sweep \\
Agent Status & Manager $\rightarrow$ Agent heartbeat & Default 40\,s \\
Session Hang & Per-state allowed time & PREPARING: 1\,h, TERMINATING: 30\,min \\
GPU Hardware & PCI bus enumeration (\texttt{lspci}) & Rev \texttt{ff}/\texttt{00} = faulty \\
GPU Metrics & \texttt{all-smi} Prometheus endpoint & Thresholds in Alertmanager \\
\bottomrule
\end{tabular}}%
\end{table}

\subsection{Unified Storage Architecture}
\label{sec:storage}

\begin{figure}[H]
  \centering
  \includegraphics[width=0.78\linewidth]{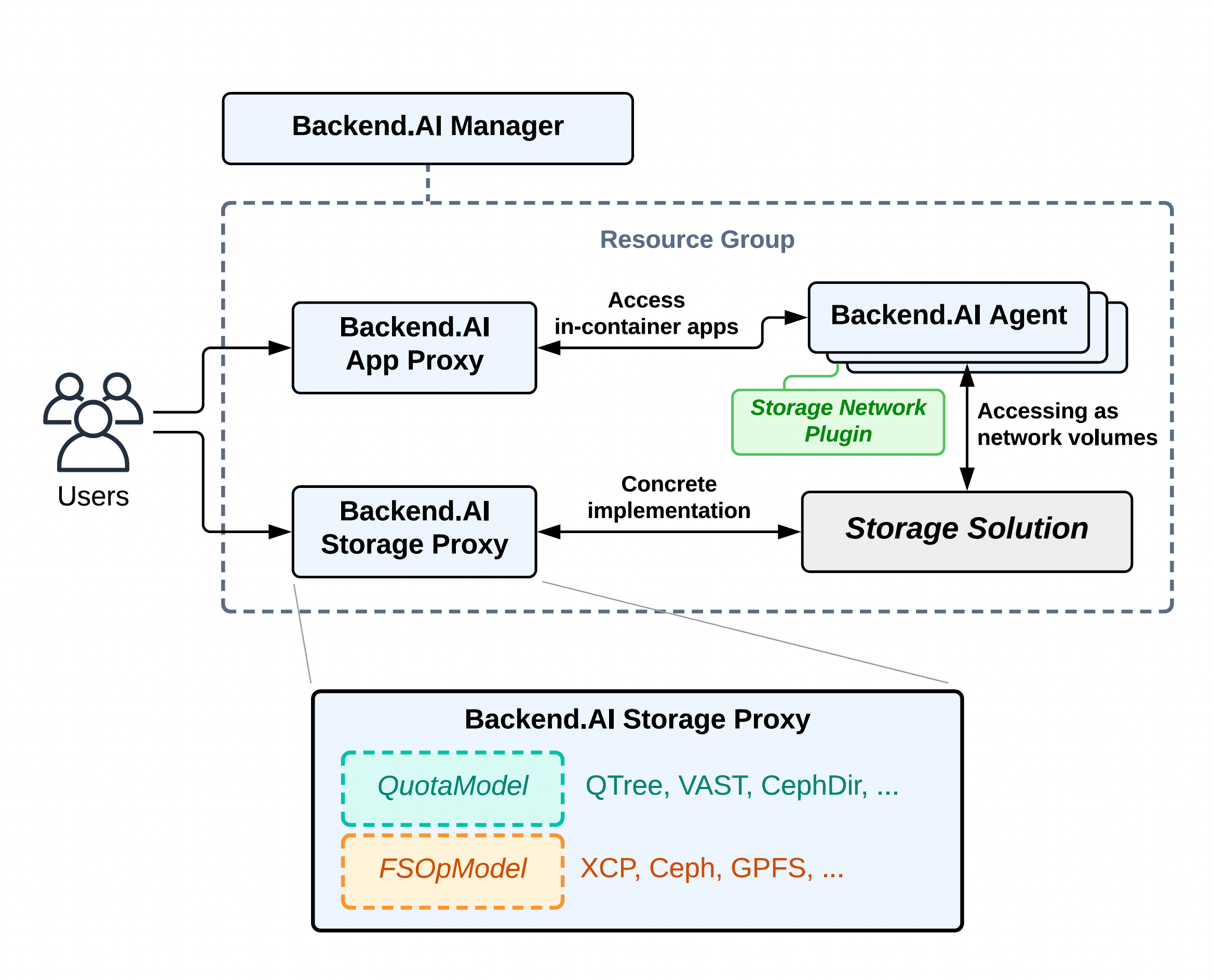}
  \caption{Backend.AI storage architecture and proxy-based integration.
  Storage resources are exposed to sessions as network volumes, with quota enforcement and filesystem operations managed by a system-level model.}
  \label{fig:storage-proxy}
\end{figure}

Backend.AI integrates storage into the session lifecycle through a proxy-based abstraction, uniformly exposing diverse storage backends (NFS, Ceph, cloud storage, etc.) as network volumes while applying quota and operational policies (Figure~\ref{fig:storage-proxy}).

\section{Glossary}

\begin{longtable}{@{}lp{10cm}@{}}
\caption{Terminology definitions}
\label{tab:terminology} \\
\toprule
\textbf{Term} & \textbf{Definition} \\
\midrule
\endfirsthead
\toprule
\textbf{Term} & \textbf{Definition} \\
\midrule
\endhead
\bottomrule
\endfoot
Agent & Backend.AI's per-node component that directly manages containers or virtual machines, allocates physical resources (GPU, CPU, memory), and reports status to the Manager \\
AOC & Active Optical Cable; an optical cable with optical-to-electrical conversion circuitry integrated at both ends. Used for longer node-to-switch links in GPU clusters where passive copper (DAC) length limits do not suffice \\
Auto-retry & Backend.AI FastTrack's automated failure recovery mechanism that detects session failures and restarts sessions so the training framework can resume from the last checkpoint; supports configurable retry count and delay \\
cuDNN & CUDA Deep Neural Network Library; NVIDIA's GPU-accelerated library of primitives for deep neural networks including convolution, normalization, and attention operations \\
DCGM & Data Center GPU Manager; NVIDIA's suite of tools for monitoring and managing GPUs in cluster environments \\
HGX & A hardware architecture that serves as NVIDIA’s data center GPU server reference platform, combining multiple GPUs with NVLink/NVSwitch. This cluster comprises 63 HGX B200 nodes \\
ECC & Error-Correcting Code; a memory protection mechanism that detects and corrects bit errors. GPU ECC errors indicate hardware-level memory defects \\
etcd & A distributed key-value store used for service discovery and configuration management in cluster systems \\
FastTrack & Backend.AI's MLOps orchestration layer that provides automated training management through auto-retry, session monitoring, and failure recovery workflows \\
FP8 / MXFP8 & 8-bit floating-point precision formats for training. FP8 reduces memory and compute costs; MXFP8 (Microscaling FP8) adds per-block scaling factors to improve numerical range \\
FSDP & Fully Sharded Data Parallel; a PyTorch distributed training strategy that shards model parameters, gradients, and optimizer states across workers \\
Gang Scheduling & All-or-nothing scheduling: either all required resources are allocated at once, or none are allocated at all \\
Goodput & The amount of useful training work completed per unit time, excluding overhead from checkpointing, communication latency, failure recovery, etc. \\
GSP & GPU System Processor; a RISC-V microcontroller running GPU firmware that communicates with the host driver via RPC. RPC timeouts are reported as XID 119 \\
HBM & High Bandwidth Memory; stacked DRAM providing high-bandwidth, high-capacity memory for GPUs. HBM3e is the variant used in NVIDIA B200 GPUs (192\,GB per device) \\
HDFS & Hadoop Distributed File System; the large-scale distributed file system of the Hadoop ecosystem. Block-based with a write-once-read-many model, optimized for big-data batch workloads \\
HSDP & Hybrid Sharded Data Parallel; a distributed training strategy combining FSDP sharding within node groups with data parallelism across groups \\
InfiniBand & A high-speed, low-latency interconnect fabric used for inter-node GPU communication in HPC and AI clusters. NDR denotes the 400\,Gbps generation \\
IOPS & Input/Output Operations Per Second; a measure of storage system throughput for random access workloads \\
IQR & Interquartile Range; the range between the 25\textsuperscript{th} and 75\textsuperscript{th} percentiles, representing the middle 50\% of a distribution \\
JCT & Job Completion Time \\
Manager & Backend.AI's central control component that coordinates cluster-wide scheduling decisions, manages session lifecycle state, and communicates with Agents via RPC \\
MFU & Model FLOPs Utilization; the ratio of observed throughput to the hardware's theoretical maximum FLOPS \\
MoE & Mixture of Experts; a model architecture that activates only a subset of parameters per input token, enabling larger total model capacity at lower per-token compute cost \\
MTBF & Mean Time Between Failures \\
NCCL & NVIDIA Collective Communications Library; provides optimized collective operations (all-reduce, broadcast, etc.) for multi-GPU and multi-node training \\
NFS & Network File System; a distributed file system protocol enabling shared storage access across cluster nodes \\
NIC & Network Interface Card. This cluster provisions per-node InfiniBand NICs for GPU communication and separate RoCE NICs for storage traffic \\
NUMA & Non-Uniform Memory Access; a memory architecture in multi-socket systems where memory access latency varies depending on the relative position of CPU and memory \\
NVLink & NVIDIA's high-bandwidth interconnect for direct GPU-to-GPU communication within a node \\
OOM & Out of Memory; a runtime error that occurs when a process requests more memory (typically GPU memory) than is available \\
PCIe & Peripheral Component Interconnect Express; a high-speed serial bus connecting GPUs, NICs, and storage devices to the host \\
Prometheus & Open-source monitoring system for time-series metric collection and querying. The 751 metrics in this report were collected in Prometheus-compatible format and stored in VictoriaMetrics \\
Resource group & Backend.AI's logical partitioning unit for cluster resources. The scheduler treats each resource group independently to limit memory usage and isolate failures \\
RoCE & RDMA over Converged Ethernet; a network protocol implementing RDMA (Remote Direct Memory Access) over Ethernet. This cluster uses a dedicated 200\,Gbps RoCE NIC for storage traffic \\
RPC & Remote Procedure Call; a communication mechanism for invoking functions in another process or host. Used for Manager--Agent communication in Backend.AI and client--server requests in NFS \\
Session & A logical training job unit in Backend.AI that can span multiple containers across multiple nodes, bundling storage, configuration, and lifecycle state as a single entity \\
Sokovan & Backend.AI's orchestration layer. Integrates session scheduling (NUMA-aware placement, gang scheduling), deployment management, and route management, reacting to events via a hint-based dual loop. This report focuses on the training session scheduling functionality \\
Temporal occupancy & The fraction of the observation period during which the cluster was occupied by training sessions, calculated as cumulative session elapsed time divided by the observation period \\
XID & NVIDIA GPU error identifier; numeric codes reported by the GPU driver to classify hardware and software errors \\
\end{longtable}

\section{GPU Monitoring Dashboard}
\label{sec:appendix-dashboard}

\begin{figure}[H]
  \centering
  \includegraphics[width=\textwidth]{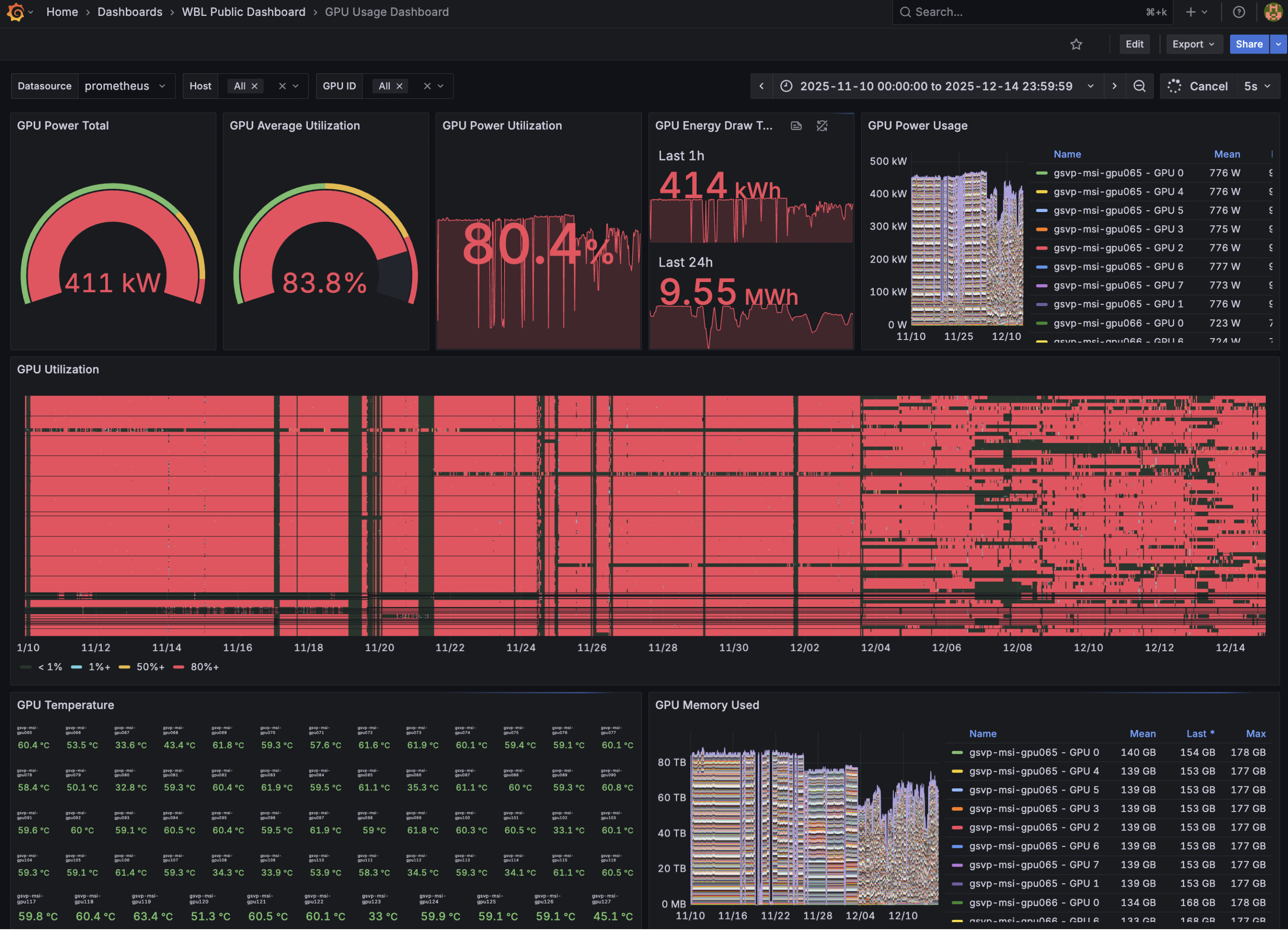}
  \caption{
    Grafana-based GPU monitoring dashboard deployed with Backend.AI.
    This dashboard provides real-time visualization of cluster-wide and per-node GPU power consumption,
    utilization, temperature, SM clocks, memory usage, and energy consumption,
    aggregating NVIDIA DCGM and \texttt{all-smi} metrics through Prometheus.
    This telemetry forms the observational basis for the failure analysis presented in Section~\ref{sec:case-studies}.
  }
  \label{fig:gpu-monitoring-dashboard}
\end{figure}

\section{Author List}\label{sec:authors}

The following is a list of authors who contributed to the development and operation of Backend.AI and the infrastructure described in this report. Names are listed alphabetically by given name.

\begin{sloppypar}
\noindent
Daemyung Kang,
Eunjin Hwang,
Hanjeong Lee\textsuperscript{*},
HyeokJin Kim,
Hyunhoi Koo,
Jeongkyu Shin,
Jeongseok Kang,
Jihyun Kang,
Jinho Heo,
Joongi Kim,
Junbum Lee,
Jungseung Yang,
Kyujin Cho,
Youngsook Song
\end{sloppypar}

\noindent\textsuperscript{*}Work done during internship at Lablup Inc.

\end{document}